\documentclass[12pt]{article} 

\usepackage{amsfonts}
\usepackage{amsmath}
\usepackage{amsthm}
\usepackage{amssymb}
\usepackage{latexsym} 
\usepackage{graphicx}
\usepackage{booktabs}
\usepackage{graphics}
\usepackage{color}
\usepackage{float}
\usepackage{hyperref}
\usepackage{adjustbox}
\usepackage[utf8]{inputenc}

\numberwithin{equation}{section}

\newcommand{\Z}{\mathbb Z}

\newcommand{\T}{\mathbb T}
\newcommand{\C}{\mathbb C}

\begin{document}
\sloppy


\begin{center}
{\Large \bf{4d $\mathcal{N}=1$ quiver gauge theories\\ and the
 $\mathrm{A_n}$ Bailey lemma}}

\vspace{1cm}

{\large \sf Frederic Br\"unner$^{a}$ and  Vyacheslav P. Spiridonov$^{b,c}$
}

\vspace{0.5cm}

\begin{itemize}
\item[$^a$] {\it  $ \makebox[-0.8em]{}$
Institut f\"ur Theoretische Physik, Technische Universit\"at Wien,\\ Wiedner Hauptstrasse 8-10, A-1040 Vienna, Austria}

\item[$^b$] $\makebox[-0.7em]{}$
{\it Laboratory of Theoretical Physics, JINR, Dubna,\\ Moscow region, 141980, Russia}

\item[$^c$] $\makebox[-0.7em]{}$
{\it National Research University Higher School of Economics, Moscow, Russia}

\end{itemize}
\end{center}

\vspace{0.5cm}

\begin{abstract}
We study the integral Bailey lemma associated with the $\mathrm{A_n}$-root system
and identities for elliptic hypergeometric integrals generated thereby. Interpreting
integrals as superconformal indices of four-dimensional $\mathcal{N}=1$ 
quiver gauge theories with the gauge groups being products of $\mathrm{SU(n+1)}$, 
we provide evidence for various new dualities. Further confirmation is achieved
by explicitly checking that the `t Hooft anomaly matching conditions holds. 
We discuss a flavour symmetry breaking phenomenon for supersymmetric quantum 
chromodynamics (SQCD), and by making use of the Bailey lemma we indicate its 
manifestation in a web of linear quivers dual to SQCD that exhibits full s-confinement.
\end{abstract}

\newpage
\tableofcontents
\newpage
\section{Introduction}

The past two decades have witnessed remarkable development in two fields
that at first glance appear to be completely unrelated: supersymmetric field theories in various space-time geometries and the theory of special functions.

The discovery of dualities in supersymmetric gauge theories has led to a new way of understanding quantum field theory at the nonperturbative level: a theory that is strongly coupled and hence not accessible with perturbative methods can have an equivalent, dual description in terms of a weakly coupled theory with entirely different degrees of freedom. The prototypical example, nowadays known as Seiberg duality \cite{Seiberg}, relates two distinct theories at an infrared conformal fixed point, one being weakly, the other strongly coupled. The theories possess $\mathcal{N}=1$ supersymmetry, and, as described in chapter 3, are cousins of quantum chromodynamics, and are therefore referred to as supersymmetric QCD (SQCD). Many other dualities for theories with various degrees of supersymmetry have been discovered, some close, some further away on the family tree of supersymmetric gauge theories. In many cases these dualities can be understood from the point of view of symmetries of higher-dimensional theories.

The requirement for two theories to be dual to each other leads to important constraints on the observables of both theories.  An important check of validity that Seiberg's example passes is the equality of triangle anomalies corresponding to global symmetries on both sides of the duality, also known as `t Hooft anomaly matching. Another observable that matches in general for dual theories is the supersymmetric partition function. For theories placed on supersymmetry-preserving background geometries, it can be shown to be equivalent \cite{CDFK} to the superconformal index \cite{KMMR,R}, a topological quantity counting BPS states which we will describe in more detail in Section 3.

Almost simultaneously the theory of special functions has independently seen important advances in the field of elliptic hypergeometric functions. One of the authors of the present article has discovered the elliptic hypergeometric integrals \cite{spi:beta}, which are at the top of the hierarchy of hypergeometric functions in the sense that they generalize plain hypergeometric functions and their q-analogs by an additional deformation parameter. Many interesting results have since been derived on the theory of these objects; see \cite{spi:essays} for an exposition to the subject. Of particular interest is the fact that there is a large number of identities relating integrals which at first glance might look rather different from each other. Many such identities have found applications in the context supersymmetric dualities as described below.

These fields have surprisingly made first contact in the work of Dolan and Osborn \cite{DO}, who have shown that the superconformal index (SCI), a quantity that as mentioned above is supposed to be equal on both sides of a duality, can actually be rewritten as an elliptic hypergeometric integral. This has opened up the possibility of approaching the study of supersymmetric partition functions from a novel point of view. As the field content on both sides of the duality is different, the corresponding integrals will also look different. Fortunately the theory of elliptic hypergeometric integrals allows for exact proofs of these equalities. In the context of $\mathcal{N}=1$ Seiberg duality, a relation known as the elliptic beta integral was used by Dolan and Osborn to prove the equivalence of the corresponding superconformal indices. Furthermore, known physical dualities led to new conjectures about unproven identities for elliptic hypergeometric integrals, leading to a fruitful interchange of mathematics and 
physics. Many explicit examples dealing with dual field theories can be found in \cite{SV1}.

Considering the above, one might ask whether it would have been possible to reverse the order of discoveries and predict Seiberg duality, starting from the corresponding integral identity and interpreting both sides as indices of supersymmetric gauge theories. Clearly the answer is yes. While it is a priori not clear if a given elliptic hypergeometric integral corresponds to an index or not, one may certainly try and examine known integral relations in the hope of uncovering evidence for dualities. This is precisely the approach we adopt in the present work.

This article deals with a technique from the theory of hypergeometric functions that, albeit well established in mathematics, has so far not seen much application in the context of physics. Starting from a simple known seed identity, the Bailey lemma allows one to generate infinite trees or chains of identities, e.g. for hypergeometric series or $q$-series. A concrete example are the Rogers-Ramanujan identities \cite{AAR}. In \cite{spi:tmf2004} it was generalized to single-variable elliptic hypergeometric integrals, which was the very first application of such ideas to integrals (as it became known
from a historical investigation in \cite{Zudilin}, Bailey himself was trying to develop the formalism
to integrals, but he could not find a use of that). A further extension of the Bailey lemma 
to elliptic hypergeometric integrals on root systems was realized in \cite{spi-war:inv}. Remarkably, the Bailey lemma also generates the star-triangle relation, allowing for the construction of solutions of the Yang-Baxter equation of the highest known level of complexity \cite{DS1}. Combined with integral identities that lie out of reach for the Bailey lemma, the tree grows even larger.

We will start by discussing the so-called ``Bailey pair" corresponding to a seed integral identity on the $\mathrm{A_n}$ root system, an elliptic hypergeometric integral that was suggested in \cite{spi:aa2003}. We consider the corresponding Bailey lemma and its applications to derivation of integral identities. A notable characteristic of the resulting Bailey tree is that it relates expressions with different numbers of integrals to each other. In particular, it gives an identity that relates an expression with an arbitrarily high number of integrals to one with none at all.

While of mathematical interest in its own right, the goal of this article is to go further and give the $\mathrm{A_n}$ Bailey tree a physical interpretation. The seed identity \eqref{An-int} itself can be understood as the equality of the superconformal indices of electric and magnetic SQCD in the case of s-confinement, where $N_f=N_c+1$ for $N_c=n+1$. The absence of an integral on the right-hand (magnetic) side indicates that the gauge group becomes trivial, in accordance with what happens in the case of s-confinement, namely that there exists a description purely in terms of gauge invariant degrees of freedom. It is therefore reasonable to assume that the integrals generated by the Bailey lemma also correspond to superconformal indices of supersymmetric gauge theories, and that the different numbers of integrals in an identity can be explained by s-confinement.

This is precisely what we find. We interpret the integral identities as indication for duality relations and after writing down the field content including $\mathrm{R}$-charges, we reinforce this evidence by explicitly confirming the 't Hooft anomaly matching conditions. In fact, the expressions containing more than one integration are superconformal indices of $\mathcal{N}=1$ linear quiver gauge theories, where nodes correspond to vector multiplets of an $\mathrm{SU(N_c)}$ gauge symmetry, and relations between quivers of different length appear to be induced by s-confinement. As shown later, the flavour symmetry structure of the Bailey quivers is nontrivial. In fact, we find that the flavour symmetries of two dual theories need not agree, indicating a symmetry breaking phenomenon that can be explained by the presence of a superpotential. We show this explicitly for a simple instance of the tree involving only one integral on each side of the duality. We also show that the Bailey tree of quivers is extended further by ordinary Seiberg duality and another set of known $\mathrm{SU}\leftrightarrow \mathrm{SU}$ identities. Some of our results were already announced in \cite{BS}.

Superconformal indices of quiver gauge theories that exhibit $s$-confinement, and more generally Seiberg duality, have appeared before in the literature. Just like in our case, the $\mathrm{A_n}$ integral relates Seiberg dual nodes of the quivers discussed in \cite{ARRY}. Similar concepts as in the present work also arise in the context of Berkooz deconfinement \cite{Berkooz}, which can be extended to an infinite sequence of dual theories possessing product gauge groups \cite{LST}. Indices of these dualities have been studied for the lowest instances in \cite{Sudano}. Even though it was suggested in \cite{SV1} that indices of the full sequence might be generated by the Bailey lemma, this problem is still open for investigation, as the involved gauge groups are of $\mathrm{SO}$ and $\mathrm{Sp}$ type, and our paper is limited to $\mathrm{SU}$.  

The article is organized as follows: in Section 2, we thoroughly study the Bailey lemma on the $\mathrm{A_n}$ root system and identities for elliptic hypergeometric integrals following from that. In Section 3, we recall the basics of the superconformal index, building the foundation for the physical interpretation of the Bailey tree. In Section 4, we discuss the structure of the quivers generated by the Bailey tree. We also show the flavour symmetry breaking phenomenon in a simple example. In Section 5, we construct larger Bailey trees and their extensions to make the full structure apparent, and finally conclude in Section 6.

\section{The Bailey lemma for the $\mathrm{A_n}$ root system}

\subsection{The lemma}

We start by defining an operation that can be thought of as an elliptic Fourier transformation
associated with the root system $\mathrm{A_n}$. Denote as $M(t)_{wz}$ an integral
operator depending on the complex parameter $t$ and acting
on meromorphic functions of the variables $z=(z_1,\ldots,z_n)$
as follows
\begin{equation}
M(t)_{wz}f(z):=\kappa_n \int_{\mathbb{T}^n}
\frac{\prod_{j,k=1}^{n+1}\Gamma(tw_jz_k^{-1};p,q)f(z)}
{\Gamma(t^{n+1};p,q)\prod_{1\leq j<k\leq n+1}\Gamma(z_jz^{-1}_k,
z_j^{-1}z_k;p,q)}\prod_{k=1}^{n}\frac{dz_k}{2\pi\textup{i}z_k},
\label{An-EFT}\end{equation}
where $\mathbb{T}^n$ is the $n$-torus (i.e., $n$-th power of the
unit circle with positive orientation), $\prod_{j=1}^{n+1}z_j=\prod_{j=1}^{n+1}w_j=1$,
and
$$
\kappa_n=\frac{(p;p)^n(q;q)^n}{(n+1)!}.
$$
In this definition we assume Einstein's convention that repeated ``indices", like $z=(z_1,\ldots,z_n)$,
denote integration over $\mathbb{T}^n$. We use the standard infinite product (the q-Pochhammer symbol)
$$
(z;p)_\infty=\prod_{k=0}^\infty(1-zp^k), \quad |p|<1, \quad z\in \C,
$$
and the elliptic gamma function  
$$
\Gamma(z;p,q)=\prod_{j,k=0}^\infty \frac{1-z^{-1}p^{j+1}q^{k+1}}
{1-zp^jq^k}, \qquad |p|,|q|<1, \quad z\in \C^*.
$$

The definition \eqref{An-EFT} is not yet precise. It is necessary to explicitly state
the class of functions on which the action of this operator is defined and the ``natural" constraints
on the parameters. We start by assuming that functions $f(z)$ are
symmetric holomorphic  functions of $z\in\C^n$. Moreover, it is convenient to
assume
$$
f(z)=f(z_1,\ldots,z_{n+1})=f(z_{\sigma_j(1)},\ldots,z_{\sigma_j(n+1)}),
\quad \prod_{k=1}^{n+1} z_k=1,
$$
where $\sigma_j\in S_{n+1}$ is any element of the permutation group of $n+1$ elements.

Next, in the original definition of $M(t)$ we impose the constraint $|tw_j|<1,\, j=1,\ldots, n+1.$
In this case the poles of the integrand in $M(t)$ are located at the points
$$
z_k=tw_jq^ap^b, \quad j,k=1,\ldots, n,\quad a,b\in\Z_{\geq 0},
$$
which form double geometric progressions going to zero for $a,b \to\infty$ and
$$
\prod_{k=1}^n z_k= t^{-1}w_j^{-1}q^{-a}p^{-b}, \quad j=1,\ldots, n+1,\quad a,b\in\Z_{\geq 0},
$$
which go to infinity for $a,b \to\infty$. Evidently, for $|tw_j|<1,\, j=1,\ldots, n+1,$
the contours of integration $z_j\in\T$ separate these two different sequences of poles.
By the Cauchy theorem it is possible to deform the domain of integration $\T^n$ so that no pole
is crossed without changing the value of the integral. One can use analytical
continuation to define the $M$-operator action for other domains of values of the parameter
$t$ and the arguments $w_j$. In particular, the normalization constant in the
definition of $M(t)$ is chosen in such a way that it allows for the definition of $M(t)$ even
for the critical values $t^{n+1}=q^{-N}p^{-M},\, N,M\in\Z_{\geq 0}$, when
a number of poles pinch the integration contours. In this case the $M$-operator turns
into a finite-difference operator, as described in detail for $n=1$ in \cite{DS1}.

The operator \eqref{An-EFT} was introduced first for $n=1$ in \cite{spi:tmf2004}; in this case
one has the additional symmetries $M(t)_{wz^{-1}}=M(t)_{w^{-1}z}=M(t)_{wz}$.
For general $n$ it was defined in \cite{spi-war:inv}.
Following these works we shall call the functions $\beta(w,t)$ and $\alpha(z,t)$
a Bailey pair with respect to the parameter $t$, if they are related to each other as

\begin{equation}
\beta(w,t)=M(t)_{wz}\alpha(z,t).
\label{An_BP}
\end{equation}

\noindent Note that one can change the variables $z_k\to 1/z_k$ in the definition \eqref{An-EFT}
and write
$$
M(t)_{wz}f(z)=M(t)_{wz^{-1}}f(z^{-1}).
$$
For some restrictions on the parameters the inversion relation for the analytically
continued operator $M(t)$ has the form \cite{spi-war:inv}
\begin{equation}
M(t)_{xw}M(t^{-1})_{wz}f(z)=f(x). \quad
\label{invAn}\end{equation}
Equivalently, it can be written as
$$
[M(t)_{wz}]^{-1} = M(t^{-1})_{wz},
$$
or, for $z_k=e^{2\pi\textup{i} a_k},\, w_k=e^{2\pi\textup{i} b_k},\,
x_k=e^{2\pi\textup{i} c_k}\in \T$, as
$$
M(t)_{wz}M(t^{-1})_{zx} = \delta_{wx} :=
\frac{1}{n!}\sum_{\sigma_j\in S_n} \prod_{k=1}^n\delta(b_k-c_{\sigma_j(k)})),
$$
where $\delta(x)$ is the Dirac delta-function. Symbolically, $M(t)M(t^{-1})=1$.

Warnaar and the second author have generalized the Bailey lemma of
\cite{spi:tmf2004} to the $\mathrm{A_n}$-root system \cite{spi-war:inv}. Let us recast it in
the notation of \cite{DS1} which we use in the following. Suppose that $\alpha(z,t)$ and
$\beta(z,t)$ form an $\mathrm{A_n}$-integral Bailey pair with respect to the
parameter $t$. Then the $\mathrm{A_n}$-Bailey lemma states that the functions
\begin{eqnarray} \label{BP_gen1} &&
\alpha'(w,st)=D(s,t^{\frac{1-n}{2}}u)_w\alpha(w,t), \quad
\\ &&
\beta'(w,st)=D(t^{-1},s^{\frac{n-1}{2}}u)_wM(s)_{wz}D(st,u)_z\beta(z,t),
\label{BP_gen2}\end{eqnarray}
where
\begin{eqnarray*} &&
D(s,u)_z:=\prod_{j=1}^{n+1}\Gamma(\sqrt{pq}s^{-\frac{n+1}{2}}uz_j^{-1},
\sqrt{pq}s^{-\frac{n+1}{2}}u^{-1}z_j;p,q),
\end{eqnarray*}
form a new Bailey pair with respect to the parameter $st$.
One can easily check that
\begin{eqnarray*} &&
D(s,u)_z=D(s,u^{-1})_{z^{-1}}, \qquad D(1,u)_z=1.
\end{eqnarray*}

For clarity we present its proof, which is completely analogous to the $n=1$ case
of \cite{spi:tmf2004}. On the one hand, we have
$$
\beta'(w,st)=D(t^{-1},s^{\frac{n-1}{2}}u)_wM(s)_{wz}D(ts,u)_zM(t)_{zx}\alpha(x,t),
$$
where we used the definition of the Bailey pairs for the function $\beta(z,t)$.
On the other hand, we should have
$$
\beta'(w,st)=M(st)_{wx}\alpha'(x,st)=M(st)_{wx}D(s,t^{\frac{1-n}{2}}u)_x\alpha(x,t).
$$
These two expressions will be equal for arbitrary functions $\alpha(x,t)$
if the following operator identity is true
$$
D(t^{-1},s^{\frac{n-1}{2}}u)_wM(s)_{wz}D(st,u)_zM(t)_{zx}
=M(st)_{wx}D(s,t^{\frac{1-n}{2}}u)_x.
$$
From the inversion relation for the elliptic gamma function one finds
that
$$
D(s,u)_zD(s^{-1},u)_z=1.
$$
Therefore we can flip the $D$-function from the left-hand side to the right and obtain
the star-triangle relation
\begin{equation}
M(s)_{wz}D(st,u)_zM(t)_{zx}=
D(t,s^{\frac{n-1}{2}}u)_wM(st)_{wx}D(s,t^{\frac{1-n}{2}}u)_x.
\label{An-STR}\end{equation}
After the substitution of explicit expressions for the operators, change of the integration
orders on the left-hand side (i.e. integrating first over the   ``internal"
variable $z$), one can see that the relation \eqref{An-STR} is true due to
the following $\mathrm{A_n}$-elliptic beta integral evaluation formula proposed in \cite{spi:aa2003}
\begin{equation}\label{An-int}
\kappa_n\int_{\mathbb{T}^n}
\frac{
\prod_{j=1}^{n+1}\prod_{k=1}^{n+2}\Gamma(s_kz_j,t_kz_j^{-1};p,q)
}{
\prod_{1\leq j< k \leq n+1}\Gamma(z_jz_k^{-1},z_j^{-1}z_k;p,q)
}\prod_{j=1}^n\frac{dz_j}{2\pi\textup{i}z_j}
=\prod_{k=1}^{n+2}\Gamma\Big(\frac{S}{s_k},\frac{T}{t_k};p,q\Big)
\prod_{k,l=1}^{n+2}\Gamma(s_kt_l;p,q).
\end{equation}
Here $S=\prod_{j=1}^{n+2}s_j$, $T=\prod_{j=1}^{n+2}t_j$, $|t_i|, |s_i|<1$ and
$ST=pq$. To deduce \eqref{An-STR} from \eqref{An-int} one has to choose
$$
t_j=tx_j^{-1}, \quad s_j=sw_j, \quad j=1,\ldots,n+1,
\quad t_{n+2}=\sqrt{pq}(st)^{-\frac{n+1}{2}}u^{-1},
\quad s_{n+2}=\sqrt{pq}(st)^{-\frac{n+1}{2}}u.
$$
For different rigorous proofs of the integration formula \eqref{An-int}
see \cite{rains,spi:rama2007}. The conditions $|s_k|, |t_k|<1$ impose
constraints upon the parameters $s, t, x_j, w_j$, which can be partially
relaxed by a change of the integration domain $\T^n\to \Omega$, such that
no singularity is crossed over during this deformation.

Let us now describe several explicit Bailey pairs generated by the
$\mathrm{A_n}$-Bailey lemma.  For this  we first apply the operator $M(t)_{wz}$ to
the Diract delta-function $\alpha(z)=\delta_{zx}$. As a result,
$$
\beta(w,t)=M(t)_{wz}\delta_{zx}=\gamma(x)\,
\frac{\prod_{j,k=1}^{n+1}\Gamma(tw_jx_k;p,q)}{\Gamma(t^{n+1};p,q)}=:M(t;w,x),
$$
where
$$
\gamma(x):=\frac{\kappa_n}
{\prod_{1\leq j<k\leq n+1}\Gamma(x_jx^{-1}_k,x_j^{-1}x_k;p,q)}.
$$
This is an example of a trivial Bailey pair, since any integral transform yields
a trivial answer after application to the delta function. Now we apply the Bailey
lemma and find
\begin{eqnarray} &&
\alpha'(w,st)=D(s,t^{\frac{1-n}{2}}u)_w\,\delta_{wx}, \quad
\\ &&
\beta'(w,st)=D(t^{-1},s^{\frac{n-1}{2}}u)_wM(s)_{wz}D(st,u)_zM(t;z,x).
\end{eqnarray}
The relation $\beta'(w,st)=M(st)_{wz}\alpha'(z,st)$ takes the form
\begin{equation}
D(t^{-1},s^{\frac{n-1}{2}}u)_wM(s)_{wz}D(st,u)_zM(t;z,x)
=M(st;w,x)D(s,t^{\frac{1-n}{2}}u)_x
\label{An-int_op}\end{equation}
and yields the left- and right-hand sides of the elliptic beta integral \eqref{An-int}
after dividing by $D(t^{-1},s^{\frac{n-1}{2}}u)_w$. However, we do not obtain a proof
of the explicit integral evaluation formula, since it was used already for proving
the Bailey lemma itself.

From \eqref{An-int_op} we immediately get the following Bailey pair:
$$
\alpha(z,s)=D(st,u)_{z}M(t;z,x), \qquad
\beta(w,s)=D(t,s^{\frac{n-1}{2}}u)_w M(st;w,x) D(s,t^{\frac{1-n}{2}}u)_x.
$$
Note that the factor $\gamma(x)$ can be cancelled from both $\alpha(z,s)$ and $\beta(w,s)$.
Replacing $s\to r,\,  u\to y$ in the next step along the stair
of Bailey pairs we obtain
\begin{eqnarray*} &&
\alpha'(z,{r}s)=D(r,s^{\frac{1-n}{2}}y)_z D(st,u)_{z}M(t;z,x), \quad
\\ &&
\beta'(w,rs)=D(s^{-1},r^{\frac{n-1}{2}}y)_wM({r})_{wz}D(rs,y)_z
D(t,s^{\frac{n-1}{2}}u)_z M(st;z,x) D(s,t^{\frac{1-n}{2}}u)_x.
\end{eqnarray*}
The relation $\beta'(w,{r}s)=M({r}s)_{wz}\alpha'(z,{r}s)$ yields a nontrivial
symmetry transformation for a particular elliptic hypergeometric integral:
\begin{eqnarray}  \label{An_id1} &&
M({r}s)_{wz}D({r},s^{\frac{1-n}{2}}y)_z D(st,u)_{z}M(t;z,x)
\\ &&
=D(s^{-1},r^{\frac{n-1}{2}}y)_w D(s,t^{\frac{1-n}{2}}u)_x M({r})_{wz}D({r}s,y)_z
D(t,s^{\frac{n-1}{2}}u)_z M(st;z,x).
\nonumber \end{eqnarray}

Let us rewrite equation \eqref{An-int_op} in the form
\begin{eqnarray} \nonumber &&
\int_{\T^n}D(t^{-1},s^{\frac{n-1}{2}}u)_wM(s;w,z)D(ts,u)_z\, \frac{\prod_{j,l=1}^{n+1}
\Gamma(tz_jx_l^{-1};p,q)}{\Gamma(t^{n+1};p,q)}\prod_{k=1}^n\frac{dz_k}{2\pi\textup{i} z_k}
 \\ &&
=D(s,t^{\frac{1-n}{2}}u)_x\frac{\prod_{j,l=1}^{n+1}
\Gamma(stw_jx_l^{-1};p,q)}{\Gamma((st)^{n+1};p,q)}.
\label{BP_gen}\end{eqnarray}
It can be iterated to yield the following explicitly computable
multiple integral
\begin{eqnarray}\label{sconfquiver}\nonumber &&
\int_{\T^{mn}}
\prod_{k=1}^m  D\left(t^{-1}S_{k-1}^{-1},s_k^{\frac{n-1}{2}}u_k\right)_{z^{(k+1)}}
M\left(s_k;z^{(k+1)},z^{(k)}\right)
D\left(tS_k,u_k\right)_{z^{(k)}}
\\  \label{s-conf-quiver}   && \makebox[6em]{} \times
\frac{\prod_{j,l=1}^{n+1}\Gamma(tz_j^{(1)}x_l^{-1};p,q)}{\Gamma(t^{n+1};p,q)}
\prod_{k=1}^m\prod_{j=1}^n\frac{dz^{(k)}_j}{2\pi\textup{i}z^{(k)}_j}
\\  &&  \makebox[4em]{}
=\prod_{k=1}^m D(s_k,\left(tS_{k-1}\right)^{\frac{1-n}{2}}u_k)_x\,
\frac{\prod_{j,l=1}^{n+1}\Gamma(tS_m z_j^{(m+1)}x_l^{-1};p,q)}
{\Gamma((tS_m)^{n+1};p,q)},
\nonumber\end{eqnarray}
where $S_k=\prod_{l=1}^ks_l$, $S_0=1$, and $u_1=u$.

\subsection{A recursion relation}

In this section we want to discuss a recursion relation arising from the Bailey lemma whose physical interpretation will form the main part of this article. 

It is convenient to define general elliptic hypergeometric integrals
(of type I) on the root system $\mathrm{A_n}$ as follows:
\begin{align} \label{An-int-def}
&I_{n}^{(m)}(s_1,\ldots,s_{n+m+2};t_1,\ldots,t_{n+m+2})=\\ 
 &\kappa_{n}\int_{\T^n} \prod_{1\leq j<k\leq n+1}
\frac{1}{\Gamma(z_jz_k^{-1},z_j^{-1}z_k;p,q)}\prod_{j=1}^{n+1}
\prod_{l=1}^{n+m+2}\Gamma(s_lz_j^{-1},t_lz_j;p,q)\; \prod_{k=1}^n
\frac{dz_k}{2\pi\textup{i}z_k},
\nonumber\end{align}
where $|t_j|, |s_j|<1$,
$$
\prod_{j=1}^{n+1}z_j=1,\qquad \prod_{l=1}^{n+m+2}s_lt_l=(pq)^{m+1},
$$
together with the convention
$$
I_{0}^{(m)}(s_1,\ldots,s_{m+2};t_1,\ldots,t_{m+2})=\prod_{l=1}^{m+2}\Gamma(s_l,t_l;p,q).
$$





As a consequence of Eq. \eqref{An-int}, the following relation for the $I_{n}^{(m)}$-integrals is valid:
\begin{align}\label{recursion1} \nonumber
& I_{n}^{(m+1)}(s_1,\ldots,s_{n+m+3};t_1,\ldots,t_{n+m+3})=\nonumber\\
&\frac{\kappa_{n}}{\Gamma(v^{n+1};p,q)}
\prod_{l=1}^{n+2}\frac{\Gamma(t_{n+m+3}s_l;p,q)}{\Gamma(v^{-n-1}t_{n+m+3}s_l;p,q)}
\int_{\T^n} \prod_{1\leq j<k\leq n+1}
\frac{1}{\Gamma(w_jw_k^{-1},w_j^{-1}w_k;p,q)}\nonumber\\
&\times
\prod_{j=1}^{n+1}
\Gamma(v^{-n}t_{n+m+3}w_j;p,q)\prod_{l=1}^{n+2} \Gamma(v^{-1}s_lw_j^{-1};p,q)\nonumber\\
&\times
I_{n}^{(m)}(vw_1,\ldots,vw_{n+1},s_{n+3},\ldots,s_{n+m+3};
t_1,\ldots,t_{n+m+2})\;\prod_{k=1}^n \frac{dw_k}{2\pi\textup{i}w_k},
\end{align}
with
$$
v^{n+1}=\frac{t_{n+m+3}}{pq}\prod_{k=1}^{n+2}s_k
=\frac{(pq)^{m+1}}{\prod_{k=1}^{n+m+2}t_k\prod_{l=n+3}^{n+m+3}s_l}.
$$

This may be checked explicitly by applying \eqref{An-int} to the integral over the $w_i$ variables. There is an analogous relation that may be obtained by changing variables as $z_i\rightarrow z_i^{-1}$ in \ref{An-int-def}. As a consequence the action on the fugacities $s_i$ and $t_i$ is exchanged, and the resulting identity is given by 

\begin{align}\label{recursion2} \nonumber
&I_{n}^{(m+1)}(s_1,\ldots,s_{n+m+3};t_1,\ldots,t_{n+m+3})=\\   
&\frac{\kappa_{n}}{\Gamma(\tilde{v}^{n+1};p,q)}
\prod_{l=1}^{n+2}\frac{\Gamma(s_{n+m+3}t_l;p,q)}{\Gamma(\tilde{v}^{-n-1}s_{n+m+3}t_l;p,q)}
\int_{\T^n} \prod_{1\leq j<k\leq n+1}
\frac{1}{\Gamma(w_jw_k^{-1},w_j^{-1}w_k;p,q)}\nonumber\\  
&\times\prod_{j=1}^{n+1}\Gamma(\tilde{v}^{-n}s_{n+m+3}w_j;p,q)\prod_{l=1}^{n+2} \Gamma(\tilde{v}^{-1}t_lw_j^{-1};p,q)\nonumber\\ 
&\times I_{n}^{(m)}(s_1,\ldots,s_{n+m+2};\tilde{v}w_1,\ldots,\tilde{v}w_{n+1},t_{n+3},\ldots,t_{n+m+3})\;\prod_{k=1}^n \frac{dw_k}{2\pi\textup{i}w_k},
\end{align}
with
$$
\tilde{v}^{n+1}=\frac{s_{n+m+3}}{pq}\prod_{k=1}^{n+2}t_k
=\frac{(pq)^{m+1}}{\prod_{k=1}^{n+m+2}s_k\prod_{l=n+3}^{n+m+3}t_l}.
$$

Eqs. (\ref{recursion1}) and (\ref{recursion2}) may be viewed as recursion relations on the set of $\mathrm{A_n}$ integrals that increase both the number of $s_i$ and $t_i$ variables by $1$ in each iteration. Note that there is no condition for this chain to ever end, i.e. one may go to arbitrarily large $m$. Note that the choice of a subset of $n+1$ parameters on which to act with this iterative operation is arbitrary in both cases. This implies that there are in total $2\times\binom{n+m+2}{n+1}$ distinct relations. Furthermore, iterations of both types may be mixed in arbitrary combinations.

In order to make notation more compact for later use, we write these general recursions arising from \eqref{recursion1} and \eqref{recursion2} respectively as 

\begin{equation}\label{r1}
 I^{(m+1)}_n=\mathcal{S}_n^{m}(\mathbf{s}) I_n^{(m)}
\end{equation}

\noindent and 

\begin{equation}\label{r2}
 I^{(m+1)}_n=\mathcal{T}_n^{m}(\mathbf{t}) I_n^{(m)},
\end{equation}

\noindent where $\mathbf{s}$ and $\mathbf{t}$ denote the sets of parameters on which the integral operators act.

Let us consider the simplest example that may arise from these relations. Substituting $m=0$ replacing $I_{n}^{(0)}$ on the right-hand side of Eq. (\ref{recursion1})
by its explicit expression, we arrive at an expression with two integrals. Applying Eq. (\ref{An-int}) to evaluate the $z_i$ integration we obtain the symmetry transformation
\begin{eqnarray}\label{enhancement} \nonumber
&&
I_{n}^{(1)}(s_1,\ldots,s_{n+3};t_1,\ldots,t_{n+3})
=\prod_{k=1}^{n+2}\Gamma\left(t_{n+3}s_k,\frac{\prod_{i=1}^{n+2}s_i}{s_k},
s_{n+3}t_k,\frac{\prod_{i=1}^{n+2}t_i}{t_k};p,q\right)
\\  && \makebox[3em]{} \times
I_{n}^{(1)}(v^{-1}s_1,\ldots,v^{-1}s_{n+2},v^{n}s_{n+3};
vt_1,\ldots,vt_{n+2}, v^{-n}t_{n+3}),
\label{m=1trafo}\end{eqnarray}
where  $\prod_{k=1}^{n+3}t_ks_k=(pq)^2$ and
$$
v^{n+1}=\frac{t_{n+3}}{pq}\prod_{k=1}^{n+2}s_k
=\frac{pq}{ s_{n+3}\prod_{k=1}^{n+2}t_k}.
$$
This relation was first derived in \cite{spi:aa2003} and it coincides with
the equality \eqref{An_id1}, which becomes evident after the identifications
\begin{eqnarray*} &&
s_j=rsw_j, \qquad t_j=tx_j^{-1}, \quad j=1,\ldots,n+1, \quad
s_{n+2}=\sqrt{pq}r^{-\frac{n+1}{2}}s^{\frac{1-n}{2}} y,
\\ &&
s_{n+3}=\sqrt{pq}(st)^{-\frac{n+1}{2}} u, \quad
t_{n+2}=\sqrt{pq}(st)^{-\frac{n+1}{2}} u^{-1}, \quad
t_{n+3}=\sqrt{pq}r^{-\frac{n+1}{2}}s^{\frac{n-1}{2}} y^{-1}
\end{eqnarray*}
and the subsequent identification $v= s$.

It is convenient to denote
\begin{eqnarray} \nonumber
&& I(v)=\prod_{k=1}^{n+2}\Gamma(v^{-n-1}s_{n+3}s_k,v^{n+1}t_{n+3}t_k;p,q)
\\ && \makebox[2em]{} \times
I_{n}^{(1)}(vs_1,\ldots,vs_{n+2}, v^{-n}t_{n+3};
v^{-1}t_1,\ldots,v^{-1}t_{n+2},v^ns_{n+3}),
\label{m=1trafo'}\end{eqnarray}
where the arguments of $I_{n}^{(1)}$ lie inside the unit circle,
$\prod_{k=1}^{n+3}t_k=\prod_{k=1}^{n+3}s_k=pq$, and $v$ is an
arbitrary free parameter (the total number of free parameters is
equal to $2n+5$). Then the derived relation can be rewritten
in the form
\begin{equation}
I(v)=I(v^{-1}).
\label{inversion}\end{equation}

\noindent Of course an analogous calculation can be done for Eq. \eqref{recursion2} with $m=0$, however, as $\tilde{v}=v^{-1}$, the result is actually identical to Eq. \eqref{enhancement}, and no new information is gained. With the idea of interpreting these identities in terms of physics in mind, it is instructive to remember that one may obtain both sides of this equation by starting with the expression on the right hand side of Eq. \eqref{recursion1} by applying Eq. \eqref{An-int} to either the $z_i$ or the $w_i$ integral.

It is important to realize that there exists a generalization of Eq. \eqref{An-int} that was found by Rains in \cite{rains}.
Denote
$$
T=\prod_{j=1}^{n+m+2}t_j,\quad S=\prod_{j=1}^{n+m+2}s_j,
$$
so that $ST=(pq)^{m+1}$, and let all $|t_k|,\, |s_k|,\, |T^{\frac{1}{m+1}}/t_k|,\,
|S^{\frac{1}{m+1}}/s_k|<1$. Then we may write
\begin{eqnarray}\nonumber\label{dual}
&&
I_{n}^{(m)}(t_1,\ldots,t_{n+m+2};s_1,\ldots,s_{n+m+2})
=\prod_{j,k=1}^{n+m+2}\Gamma(t_js_k;p,q) \\ && \makebox[1em]{} \times
I_{m}^{(n)}\left(\frac{T^{\frac{1}{m+1}}}{t_1},\ldots,
\frac{T^{\frac{1}{m+1}}}{t_{n+m+2}};
\frac{S^{\frac{1}{m+1}}}{s_1},\ldots,\frac{S^{\frac{1}{m+1}}}{s_{n+m+2}}\right).
\label{rains2}\end{eqnarray}

\noindent This relation, which provides a substantial check for Seiberg duality, may be combined for $m>0$ with the Bailey recursions defined above in an analogous manner as for $m=0$, in order to produce more complicated identities. As described later, we will use this fact to build a large tree of equivalences between superconformal indices of $\mathcal{N}=1$ supersymmetric field theories. For this purpose, let us introduce notation that may be easily combined with Eqs. \eqref{r1} and \eqref{r2}, i.e. let us rewrite Eq. \eqref{dual} simply as

\begin{equation}\label{crec}
I_n^{(m)}=\mathcal{C}_n^mI_m^{(n)}.
\end{equation}

\noindent In this language, Eq. \eqref{enhancement} is given by 

\begin{equation}
 I_n^{(1)}=\mathcal{S}_n^{0}(\mathbf{s})\mathcal{C}_n^0I_0^{(n)}.
\end{equation}

\noindent We may now also combine Eqs. \eqref{r1} and \eqref{crec} more generally as

\begin{equation}
 I^{(m+1)}_n=\mathcal{S}_n^{m}(\mathbf{s}) \mathcal{C}_n^m I_m^{(n)}.
\end{equation}

\noindent Written out explicitly for $\mathbf{s}=\{s_1,\dots s_{n+1}\}$, this corresponds to

\begin{align}\label{BS}
&I_{n}^{(m+1)}(s_1,\ldots,s_{n+m+3};t_1,\ldots,t_{n+m+3})=\nonumber\\
&\frac{\kappa_{n}\kappa_m}{\Gamma(v^{n+1};p,q)}
\prod_{l=1}^{n+2}\frac{\Gamma(t_{n+m+3}s_l;p,q)}{\Gamma(v^{-n-1}t_{n+m+3}s_l;p,q)}\prod_{k=1}^{n+m+2}\prod_{j=1}^{n+1}\Gamma(v w_j t_k;p,q) \prod_{l=1}^{m+1}(s_{l+n+2}t_k;p,q)\nonumber\\
&\times\int_{\T^n}\int_{\T^m} \prod_{1\leq j<k\leq n+1}
\frac{1}{\Gamma(w_jw_k^{-1},w_j^{-1}w_k;p,q)}\prod_{1\leq j<k\leq m+1}
\frac{1}{\Gamma(z_jz_k^{-1},z_j^{-1}z_k;p,q)}\nonumber\\
&\times \prod_{j=1}^{m+1}\prod_{l=1}^{n+1}\Gamma(S^{\frac{1}{m+1}}v^{-1}w_l^{-1}z_j^{-1};p,q)\prod_{k=1}^{m+1}\Gamma(S^{\frac{1}{m+1}}s_{k+n+2}^{-1}z_j^{-1};p,q)\prod_{i=1}^{n+m+2}\Gamma(T^{\frac{1}{m+1}}t_i^{-1}z_j;p,q)\nonumber\\
&\times\prod_{j=1}^{n+1} \Gamma(v^{-n}t_{n+m+3}w_j;p,q)\prod_{l=1}^{n+2} \Gamma(v^{-1}s_lw_j^{-1};p,q)\prod_{k=1}^n\frac{dw_k}{2\pi\textup{i}w_k}\prod_{i=1}^m\frac{dz_i}{2\pi\textup{i}z_i}.
\end{align}

\noindent An analogous relation exists also for the recursion corresponding to Eq. \eqref{r2}.

To conclude this section, we give an explicit example of an integral arising from a combination of Eqs. \eqref{r1} and \eqref{r2}. We start with the integral $I_n^{(m)}$, and first apply Eq. \eqref{r1}, leading to an expression for $I_n^{(m+1)}$ given in terms of two integrals. Then we apply Eq. \eqref{r2} acting on the other set of parameters, resulting in an expression for $I_n^{(m+2)}$ given by

\begin{align}\label{snowflake}
&I_{n}^{(m+2)}(s_1,\ldots,s_{n+m+4};t_1,\ldots,t_{n+m+4})
=
\frac{\kappa_{n}}{\Gamma(v_s^{n+1};p,q)}
\prod_{l=1}^{n+2}\frac{\Gamma(t_{n+m+3}s_l;p,q)}{\Gamma(v_s^{-n-1}t_{n+m+3}s_l;p,q)}\nonumber\\
&\times\int_{\T^n} \prod_{1\leq j<k\leq n+1}
\frac{1}{\Gamma(w_jw_k^{-1},w_j^{-1}w_k;p,q)}\prod_{j=1}^{n+1}
\Gamma(v_s^{-n}t_{n+m+3}w_j;p,q)\prod_{l=1}^{n+2} \Gamma(v_s^{-1}s_lw_j^{-1};p,q)\nonumber\\
&\times\frac{\kappa_{n}}{\Gamma(v_t^{n+1};p,q)}
\prod_{l=1}^{n+2}\frac{\Gamma(s_{n+m+4}t_l;p,q)}{\Gamma(v_t^{-n-1}s_{n+m+4}s_l;p,q)}
\int_{\T^n} \prod_{1\leq j<k\leq n+1}\frac{1}{\Gamma(y_jy_k^{-1},y_j^{-1}y_k;p,q)}\nonumber\\
&\times I_{n}^{(m)}(v_sw_1,\ldots,v_sw_{n+1},s_{n+3},\ldots,s_{n+m+3};
v_ty_1,\ldots,v_ty_{n+1},t_{n+3},\ldots,t_{n+m+2})\nonumber\\
&\times\prod_{j=1}^{n+1}
\Gamma(v_t^{-n}s_{n+m+4}w_j;p,q)\prod_{l=1}^{n+2} \Gamma(v_t^{-1}t_ly_j^{-1};p,q)\prod_{k=1}^n \frac{dw_k}{2\pi\textup{i}w_k}
\prod_{k=1}^n \frac{dy_k}{2\pi\textup{i}y_k},
\end{align}

\noindent with

\begin{equation}
v_s^{n+1}=\frac{t_{n+m+3}}{pq}\prod_{k=1}^{n+2}s_k
\end{equation}

\noindent and

\begin{equation}
v_t^{n+1}=\frac{s_{n+m+4}}{pq}\prod_{k=1}^{n+2}t_k.
\end{equation}

\section{SQCD and superconformal index}

\subsection{The index}


The superconformal index (SCI) \cite{KMMR,R} is quantity counting BPS states of a superconformal theory in the sense that it receives contributions exclusively from short representations that cannot be combined into long representations. It can also be thought of as a supersymmetric partition function of the theory put on a primary Hopf surface, a manifold of the same topology as $\mathrm{S^1\times S^3}$. Note that it is normalised in such a way that the ground state contribution is precisely $1$. A localization computation \cite{ACM} has shown that the actual partition function is proportional to the SCI up to a normalization factor involving the supersymmetric Casimir energy that contains the anomaly polynomial and can be computed directly in several ways \cite{BBK,MS,BRS}. The index is invariant under continuous transformations that do not violate superconformal invariance of the theory. Of particular interest is the fact that it is invariant under the deformation by marginal operators 
corresponding to the renormalization group flow. As a consequence it may be computed in a regime where the theory is free.

The explicit form of the SCI is given by a generalization of the Witten index as

\begin{equation}
\mathcal{I}(p,q,y)=\mathrm{Tr_{\mathcal{H}(S^3)}}(-1)^{\mathcal{F}}e^{-\beta H}p^{\frac{R}{2}+J_R+J_L}q^{\frac{R}{2}+J_R-J_L}\prod_{j}y_j^{F_j},
\end{equation}

\noindent where the trace is over modes living the Hilbert space on $\mathrm{S^3}$, $\mathcal{H}(S^3)$, $R$ is the $R$-charge, $J_L$ and $J_R$ are the Cartan generators of the rotation group $\mathrm{SU(2)_L\times SU(2)_R}$. 
Here $(-1)^{\mathcal{F}}$ is the $\mathbb{Z}_2$-grading operator because of which the index only receives contributions from (gauge invariant) states with $H=E-2J_L-\frac{3}{2}R=0$, $E$ being the energy. Therefore the index only depends on the fugacities $p$ and $q$ which are complex structure moduli of the background geometry, and the fugacities $y_j$ corresponding to the flavor symmetries of the theory ($F_j$ denoting the Cartan generators of respective groups), and is independent of the chemical potential $\beta$. The index can be evaluated explicitly following prescription of \cite{KMMR,R}, which tells us that it is equivalent to a gauge-invariant integral over the so-called plethystic exponential of the single-particle state (or single letter) indices $i(p^n,q^n,y^n,z^n)$, i.e.

\begin{equation}\label{index}
\mathcal{I}(p,q,y)=\int_G d\mu(g)\;\mathrm{exp}\left(\sum_{n=1}^\infty\frac{1}{n}i(p^n,q^n,y^n,z^n)\right).
\end{equation}

\noindent The integral can be interpreted as a continuous sum over all flat connections on the $\mathrm{S^1}$ part of the geometry, as made precise by supersymmetric localization \cite{ACM}.  The single-particle state index is a quantity that depends on the characters of both the gauge and the flavor groups:

\begin{equation}\label{spi}
i(p,q,y,z)=\frac{2pq-p-q}{(1-p)(1-q)}\chi_{adj}(z)+\sum_j\frac{(pq)^{R_j/2}\chi_j(y)\chi_j(z)-(pq)^{1-R_j/2}\overline{\chi}_j(y)\overline{\chi}_j(z)}{(1-p)(1-q)},
\end{equation}

\noindent where $\chi_{adj}(z)$ is the character of the adjoint representation under which the gauge fields transform, while the second term is a sum over the chiral matter superfields that contains the characters of the corresponding representations of the gauge and flavor groups. The single particle index itself is counting holomorphic sections of line bundles, and as such can be related to a certain index-character \cite{MS}.

Dolan and Osborn discovered that the integral of Eq. (\ref{index}) could be rewritten as an elliptic hypergeometric integral. The main observation is that plethystic exponentials of expressions like Eq. (\ref{spi}) can be turned into combinations of q-Pochhammer symbols corresponding to elliptic gamma functions. This can be seen from 

\begin{equation}
 \mathrm{exp}\left(\sum_{n=1}^\infty\frac{1}{n}\frac{x^n}{(1-p^n)(1-q^n)}\right)=\prod_{n=1}^\infty\mathrm{exp}\left(\frac{1}{n}\frac{x^n}{(1-p^n)(1-q^n)}\right)=(x;q,p)_\infty^{-1},
\end{equation}

\noindent where

\begin{equation}
 (x;q,p)_\infty=\prod_{i,j=0}^\infty(1-x p^i q^j)
\end{equation}

\noindent Denotes a generalized q-Pochhammer symbol. The elliptic gamma function then arises as 

\begin{equation}
 \Gamma(x;p,q)=\frac{(x^{-1}pq;p,q)_\infty}{(x;p,q)_\infty},
\end{equation}

\noindent and is defined for $x\in\mathbb{C}^*$.

As a consequence, the whole machinery of elliptic hypergeometric integrals \cite{spi:beta,spi:essays,rains}
can be applied to the study of supersymmetric dualities.

\subsection{Supersymmetric QCD}

\begin{table}[t]
\centering
\begin{tabular}{lccccc}
\toprule
 &  $\mathrm{SU(N_c)}$  & $\mathrm{SU(N_f)_s}$ & $\mathrm{SU(N_f)_t}$ & $\mathrm{U(1)_B}$ & $\mathrm{U(1)_R}$\\
\toprule
$\mathbf{Q}^i$ & $f$ & $f$ & $1$ & $1$ & $(N_f-N_c)/N_f$ \\
$\mathbf{\tilde{Q}}_i$ & $\bar{f}$  &  $1$ & $\bar{f}$ & $-1$ & $(N_f-N_c)/N_f$ \\
$\mathbf{V}$ & $\mathrm{adj}$  & $1$ & $1$ & $0$ & $1$ \\
\toprule
\end{tabular}
\caption{The matter content of electric $\mathcal{N}=1$ SQCD.}
\label{electricmatter}
\end{table}

\begin{table}[t]
\centering
\begin{tabular}{lccccc}
\toprule
 &  $\mathrm{SU(\tilde{N}_c)}$  & $\mathrm{SU(N_f)_s}$ & $\mathrm{SU(N_f)_t}$ & $\mathrm{U(1)_B}$ & $\mathrm{U(1)_R}$\\
\toprule
$\mathbf{q}^i$ & $f$ & $\bar{f}$ & $1$ & $N_c/\tilde{N_c}$ & $N_c/N_f$ \\
$\mathbf{\tilde{q}}_i$ & $\bar{f}$  &  $1$ & $f$ & $-N_c/\tilde{N_c}$ & $N_c/N_f$ \\
$\mathbf{V'}$ & $\mathrm{adj}$  & $1$ & $1$ & $0$ & $1$ \\
$\mathbf{M}$ & $1$  & $f$ & $\bar{f}$ & 0 & $\tilde{N_c}/N_f$ \\
\toprule
\end{tabular}
\caption{The matter content of magnetic $\mathcal{N}=1$ SQCD.}
\label{magneticmatter}
\end{table}

In this section, we will shortly review Seiberg duality \cite{Seiberg} for $\mathcal{N}=1$ SQCD, since   it is important for the remainder of the paper. SQCD with $\mathrm{SU(N_c)}$ gauge group and $N_f$ flavours consists of quarks and squarks in chiral multiplets $\mathbf{Q}^i$ and $\mathbf{\tilde{Q}}_i$, and a gauge field in the vector multiplet $\mathbf{V}$. This theory, which in the spirit of Montonen-Olive duality is commonly referred to as {\em electric},  has a $\mathrm{U(1)_R}$ symmetry and a $\mathrm{SU(N_f)\times SU(N_f)\times U(1)_B}$ flavour symmetry. The matter content is summarized in Table \ref{electricmatter}, where the entries denote representations under which the fields transform for $\mathrm{SU}$ symmetries, and charges for $\mathrm{U(1)}$ symmetries. Seiberg discovered that for $3N_c/2<N_f<3N_c$, there exists an infrared fixed point at which the theory is conformal. Furthermore he found that at this fixed point, there exists a dual description:
{\em magnetic} SQCD with gauge group $\mathrm{SU(\tilde{N}_c)}$ for $\tilde{N}_c=N_f-N_c$. The flavour symmetries are the same as in the electric theory. In addition to fundamental magnetic quarks and squarks in the chiral multiplets $\mathbf{q}^i$ and $\mathbf{\tilde{q}}_i$ and the vector multiplet $\mathbf{V'}$, there now exists a fundamental mesonic degree of freedom $\mathbf{M}$. The latter is coupled to the magnetic quark multiplets through a nonvanishing superpotential. The matter content of this theory can be found in Table \ref{magneticmatter}. It is noteworthy that while the superpotential of the electric theory vanishes, that of the magnetic theory is given by $W\propto \mathbf{M} \mathbf{q}^i\mathbf{\tilde{q}}^i$. Both theories are asymptotically free and flow to the aforementioned fixed point.

An interesting phenomenon known as \emph{s-confinement} arises for $N_f=N_c+1$: in this case the gauge group of the magnetic side is trivial and a confining superpotential is generated dynamically. The degrees of freedom are gauge-singlet composite states made of the fundamental fields of the theory. Note that the quarks remain massless, i.e. there is no breaking of chiral symmetry.  

An important check of Seiberg duality is that global 't Hooft anomalies as captured by triangle diagrams match for both sides of the theory. A more refined check, that also captures information about anomalies, has by now become standard technology, is to confirm that the SCIs, as defined in the previous chapter, of both descriptions match. This is to be expected, since for two dual descriptions the physical content, and hence any function counting states, should be identical. A proof of this was first achieved by Dolan and Osborn \cite{DO} by realizing that the SCIs of electric and magnetic SQCD can be rewritten as elliptic hypergeometric integrals. In fact, the SCI of the electric theory is given by Eq. (\ref{An-int-def}), as can be seen after the identification $N_c=n+1$ and $N_f=n+m+2$. The fugacities of the flavour group $\mathrm{SU(N_f)_s\times SU(N_f)_t\times U(1)_B}$ are hidden in the definition of the parameters $s_l$ and $t_l$. R-charges and baryon number can be restored by setting $s_l\rightarrow (pq)^{\frac{1}{2}R_{\mathbf{\tilde{Q}}}-x}\tilde{
t}_
l^{-1}$ and $t_l\rightarrow (pq)^{\frac{1}{2}R_{\mathbf{Q}}+x}\tilde{s}_l$, where $R_{\mathbf{Q}}$ and $R_{\mathbf{\tilde{Q}}}$ are the R-charges of the chiral quark and antiquark multiplets (given in Table \ref{electricmatter}), respectively, and $(pq)^x$ is the fugacity for $\mathrm{U(1)
_B}$, while $\tilde{s}_i$ and $\tilde{t}_i$ are the fugacities of $\mathrm{SU(N_f)_s\times SU(N_f)_t}$ satisfying $\prod_{i=1}^{N_f}\tilde{s}_i=\prod_{i=1}^{N_f}\tilde{t}_i=1$. Matching of the SCIs is proven through Eq. (\ref{dual}), where the right-hand side gives the index of the magnetic theory, which can be confirmed by changing variables as described above.

This was the first instance of a long series of successful applications of techniques for elliptic hypergeometric functions to supersymmetric gauge theories. By applying this logic to other integrals as well, we will determine the matter content of the theories whose superconformal indices are generated by the Bailey lemma, as discussed in subsequent sections. This way, we will not use the mathematical apparatus to confirm known dualities, but rather use it to predict new physical relationships.

\section{A simple Bailey quiver}

\subsection{Field content}

The phenomenon of $s$-confinement as described above is not limited to SQCD, but can be generalized to occur also in a class of supersymmetric gauge theories known as \emph{quivers}. Their field content can be summarized in so-called quiver diagrams, which are graphs consisting of nodes and edges. Nodes denote vector multiplets in the adjoint representation of distinct gauge groups, which means that quivers in general admit more than one gauge symmetry. Edges connecting two nodes stand for bifundamental multiplets, transforming in the fundamental of one group and in the antifundamental of the other. There will in general also be nodes corresponding to flavour symmetries.

In this section we study the full physical interpretation of the expressions on the right hand sides of Eqs. \eqref{recursion1} and \eqref{recursion2}. The statements about other sets of parameters, as written more generally in Eqs. \eqref{r1} and \eqref{r2}, will be identical. Their distinction will only play a role once we consider more complicated constructions in later sections.

Let us begin by rewriting the right hand side of Eq. \eqref{recursion1} in terms of the variables introduced in the previous section, suggestively denoting the expression as $\mathcal{I}_{\mathrm{quiver},\mathcal{S}}$ to indicate that we will interpret it as a superconformal index:

\begin{align}\label{quiverindex1}
&\mathcal{I}_{\mathrm{quiver},\mathcal{S}}=\kappa_n^2\Gamma\left((pq)^{1+\frac{N_c}{2N_f}(2+N_c-N_f)+N_c x}\tilde{s}_{N_f}^{-1}U\right)\nonumber\\
&\prod_{l=1}^{N_c+1}\Gamma\left((pq)^{\frac{N_f-N_c}{N_f}}\tilde{s}_{N_f}U^{-\frac{1}{N_c+1}}\tau_{1,l}^{-1}\right)\Gamma\left((pq)^{\frac{N_c(N_f-N_c)}{2N_f}-N_c x}U^{-\frac{N_c}{N_c+1}}\tau_{1,l}\right)\nonumber\\
&\int_{\mathbb{T}_w^n}\int_{\mathbb{T}_z^n}\prod_{1\leq j<k\leq N_c}
\frac{1}{\Gamma(w_jw_k^{-1},w_j^{-1}w_k,z_jz_k^{-1},z_j^{-1}z_k)}\nonumber\\
&\prod_{j=1}^{N_c}\Gamma\left((pq)^{\frac{(N_c^2-N_c N_f+2N_f-2)}{2N_f}+N_c x}\tilde{s}_{N_f}^{\frac{1}{N_c}}U^{\frac{N_c-1}{N_c}}w_j\right)\prod_{l=1}^{N_c+1}\Gamma\left((pq)^{\frac{1}{N_f}}\tilde{s}_{N_f}^{-1}U^{\frac{1}{N_c}-\frac{1}{N_c+1}}\tau_{1,l}^{-1}w_j^{-1}\right)\nonumber\\
&\prod_{r=1}^{N_c}\Gamma\left((pq)^{-\frac{N_c(2+N_c-N_f)}{2N_f}-x}\tilde{s}_{N_f}^{\frac{1}{N_c}}U^{-\frac{1}{N_c}}w_r z_j^{-1}\right)\prod_{k=1}^{N_f-N_c-1}\Gamma\left((pq)^{\frac{N_f-N_c}{2N_f}-x}U^{\frac{1}{N_f-N_c-1}}\tau_{2,k}^{-1}z_j^{-1}\right)\nonumber\\
&\prod_{i=1}^{N_f-1}\Gamma\left((pq)^{\frac{N_f-N_c}{2N_f}+x}\tilde{s}_{N_f}^{-\frac{1}{N_f-1}}\sigma_i z_j\right)\prod_{a=1}^{N_c-1}\frac{dw_a}{2\pi \textup{i} w_a}\frac{dz_a}{2\pi \textup{i} z_a},
\end{align}

\noindent where $U=\prod_{k=1}^{N_c+1}\tilde{t}_k$, $\tau_{1,l}=U^{-\frac{1}{N_c+1}}\tilde{t}_l$, $\tau_{2,l}=U^{\frac{1}{N_f-N_c-1}}\tilde{t}_{N_c+1+l}$ and $\sigma_l=\tilde{s}_{N_f}^{\frac{1}{N_f-1}}\tilde{s}_l$. 

This expression matches the generic form of a superconformal index after inserting all relevant characters into the plethystic exponential and integrating over the $\mathrm{S^1}$ flat connections of two distinct gauge groups $\mathrm{SU(N_c)_z}$ and $\mathrm{SU(N_c)_w}$ , where we have identified $\mathrm{N_c}=n+1$. We may identitfy the correct contributions of two $\mathcal{N}=1$ vector multiplets, with corresponding holonomies $z_i$ and $w_i$. Furthermore, there appears to be highly nontrivial field content that may be described as a combination of $\mathcal{N}=1$ chiral multiplets transforming in representations of various global symmetry groups.
From $\prod_{l=1}^{N_c+1}\tau_{1,l}=\prod_{l=1}^{N_f-N_c-1}\tau_{2,l}=\prod_{l=1}^{N_f-1}\sigma_l=1$, we see that the global symmetries are given, in addition to $\mathrm{U(1)_B}$ and $\mathrm{U(1)_R}$, by $\mathrm{SU(N_c+1)_{\tau_1}}$, $\mathrm{SU(N_f-N_c-1)_{\tau_2}}$, $\mathrm{ U(1)_{\tau}}$, $\mathrm{SU(N_f-1)_{\sigma}}$, and $\mathrm{U(1)_{\sigma}}$, where $U$ and $\tilde{s}_{N_f}$ act as the fugacities of $\mathrm{U(1)_{\tau}}$ and $\mathrm{U(1)_{\sigma}}$, respectively. In fact, these symmetry groups are regular subgroups of the global symmetries of SQCD, i.e. $\mathrm{SU(N_c+1)_{\tau_1}\times SU(N_f-N_c-1)_{\tau_2}\times U(1)_{\tau}}\subset\mathrm{SU(N_f)_t}$, and $\mathrm{SU(N_f-1)_\sigma\times U(1)_\sigma}\subset \mathrm{SU(N_f)_s}$. To see that the fugacities match, we decompose the superconformal index of the electric theory as

\begin{table}[t]
\centering
\begin{adjustbox}{max width=\textwidth}
\begin{tabular}{lccccccccc}
\toprule
 &  $\mathrm{SU(N_c)_z}$  & $\mathrm{SU(N_c)_w}$ & $\mathrm{SU(N_f-N_c-1)_\tau}$ & $\mathrm{SU(N_c+1)_\tau}$ & $\mathrm{SU(N_f-1)_\sigma}$ & $\mathrm{U(1)_\sigma}$ & $\mathrm{U(1)_\tau}$ & $\mathrm{U(1)_B}$ & $\mathrm{U(1)_R}$\\
\toprule

$\mathbf{Q}_1$ & $f$ & $1$ & $1$ & $1$ & $f$ & $-\frac{1}{N_f-1}$ & $0$ & $1$ & $\frac{N_f-N_c}{N_f}$\\
$\mathbf{\bar{Q}}_1$ & $\bar{f}$ & $1$ & $\bar{f}$ & $1$ & $1$ & $0$ & $\frac{1}{N_f-N_c-1}$ & $-1$ & $\frac{N_f-N_c}{N_f}$\\
$\mathbf{Q}_2$ & $1$ & $f$ & $1$ & $1$ & $1$ & $\frac{1}{N_c}$ & $\frac{N_c-1}{N_c}$ & $N_c$ & $\frac{N_c^2-N_cN_f+2N_f-2}{Nf}$\\
$\mathbf{\bar{Q}}_2$ & $1$ & $\bar{f}$ & $1$ & $\bar{f}$ & $1$ & $-\frac{1}{N_c}$ & $\frac{1}{N_c}-\frac{1}{N_c+1}$ & $0$ & $\frac{2}{N_f}$\\
$\mathbf{Q}_{12}$ & $\bar{f}$ & $f$ & $1$ & $1$ & $1$ & $\frac{1}{N_c}$ & $-\frac{1}{N_c}$ & $-1$ & $\frac{N_f-N_c-2}{N_f}$\\
$\mathbf{M}$ & $1$ & $1$ & $1$ & $\bar{f}$ & $1$ & $1$ & $-\frac{1}{N_c+1}$ & $0$ & $2\frac{N_f-N_c}{N_f}$\\
$\mathbf{B}_1$ & $1$ & $1$ & $1$ & $1$ & $1$ & $-1$ & $1$ & $N_c$ & $2+\frac{N_c(N_c+2-N_f)}{N_f}$\\
$\mathbf{B}_2$ & $1$ & $1$ & $1$ & $f$ & $1$ & $1$ & $-\frac{N_c}{N_c+1}$ & $-N_c$ & $\frac{N_c(N_f-N_c)}{N_f}$\\
$\mathbf{V}_1$ & $\mathrm{adj}$ & $1$ & $1$ & $1$ & $1$ & $0$ & $0$ & $0$ & $1$\\
$\mathbf{V}_2$ & $1$ & $\mathrm{adj}$ & $1$ & $1$ & $1$ & $0$ & $0$ & $0$ & $1$\\
\toprule
\end{tabular}
\end{adjustbox}
\caption{The matter content of the quiver gauge theory whose field content corresponds to the superconformal index $\mathcal{I}_{\mathrm{quiver},\mathcal{S}}$ given by Eq. (\ref{quiverindex1}).}
\label{quivertable}
\end{table}

\begin{align}\label{subgroups1}
\mathcal{I}_{\mathrm{eSQCD}}&=\kappa_n\int_{\mathbb{T}_z^n}\prod_{1\leq j<k\leq N_c}
\frac{1}{\Gamma(z_jz_k^{-1},z_j^{-1}z_k)}\prod_{j=1}^{N_c}\prod_{l=1}^{N_f}   \Gamma\left((pq)^{r_{\tilde{Q}}}\tilde{t}_l^{-1}z_j^{-1},(pq)^{r_Q}\tilde{s}_l z_j\right)\prod_{a=1}^{N_c-1}\frac{dz_a}{2\pi \textup{i} z_a}\nonumber\\
&=\kappa_n\int_{\mathbb{T}_z^n}\prod_{1\leq j<k\leq N_c}
\frac{1}{\Gamma(z_jz_k^{-1},z_j^{-1}z_k)}\prod_{j=1}^{N_c}\prod_{l=1}^{N_c+1}\Gamma\left((pq)^{r_{\tilde{Q}}}U^{-\frac{1}{N_c+1}}\tau_{1,l}^{-1}z_j^{-1}\right) \nonumber\\
 \times \prod_{k=1}^{N_f-N_c-1}&\Gamma\left((pq)^{r_{\tilde{Q}}}U^{\frac{1}{N_f-N_c-1}}\tau_{2,k}^{-1}z_j^{-1}\right)\prod_{i=1}^{N_f-1}\Gamma\left((pq)^{r_Q}s_{N_f}^{-\frac{1}{N_f-1}}\sigma_i z_j\right)\Gamma\left((pq)^{r_Q}s_{N_f}z_i\right)\prod_{a=1}^{N_c-1}\frac{dz_a}{2\pi \textup{i} z_a},
\end{align}

\noindent with $r_Q=\frac{1}{2}R_{\mathbf{Q}}+x$ and $r_{\tilde{Q}}=\frac{1}{2}R_{\mathbf{Q}}-x$.

Eq. (\ref{recursion1}) can be iterated, starting from ordinary SQCD with $N_f=N_c+1$, corresponding to $m=0$. This means that from this equation alone, a given superconformal index of SQCD with $N_f>N_c+1$ flavours can be rewritten in terms of $N_f-N_c-1$ distinct linear quivers, where the global symmetries of additional quivers are given by subgroups of the quiver described above. It follows that SQCD, depending on the number of colours and flavours, is possibly dual to a large number of linear quivers. However, note that in principle we may act with the operation on a different set of parameters as in Eq. \eqref{r1}, and that there is a second equation we may mix into the iteration procedure, namely \eqref{r2}. We will therefore now go on and describe the quiver whose superconformal index is given by the right hand side of Eq. \eqref{recursion2}. Explicitly rewriting it in terms of the variables in an analogous manner to Eq. \eqref{quiverindex1} we obtain

\begin{align}\label{quiverindex2}
 &\mathcal{I}_{\mathrm{quiver},\mathcal{T}}=\kappa_n^2\Gamma\left((pq)^{1+\frac{N_c}{2N_f}(2+N_c-N_f)-N_c x}\tilde{t}_{N_f}U'\right)\nonumber\\
&\prod_{l=1}^{N_c+1}\Gamma\left((pq)^{\frac{N_f-N_c}{N_f}}\tilde{t}_{N_f}^{-1}U'^{\frac{1}{N_c+1}}\sigma_{1,l}\right)\Gamma\left((pq)^{\frac{N_c(N_f-N_c)}{2N_f}+N_c x}U'^{\frac{N_c}{N_c+1}}\sigma_{1,l}^{-1}\right)\nonumber\\
&\int_{\mathbb{T}_w^n}\int_{\mathbb{T}_z^n}\prod_{1\leq j<k\leq N_c}
\frac{1}{\Gamma(w_jw_k^{-1},w_j^{-1}w_k,z_jz_k^{-1},z_j^{-1}z_k)}\nonumber\\
&\prod_{j=1}^{N_c}\Gamma\left((pq)^{\frac{(N_c^2-N_c N_f+2N_f-2)}{2N_f}-N_c x}\tilde{t}_{N_f}^{-\frac{1}{N_c}}U'^{-\frac{N_c-1}{N_c}}w_j\right)\prod_{l=1}^{N_c+1}\Gamma\left((pq)^{\frac{1}{N_f}}\tilde{t}_{N_f}U'^{-\frac{1}{N_c}+\frac{1}{N_c+1}}\sigma_{1,l}w_j^{-1}\right)\nonumber\\
&\prod_{r=1}^{N_c}\Gamma\left((pq)^{-\frac{N_c(2+N_c-N_f)}{2N_f}+x}\tilde{t}_{N_f}^{-\frac{1}{N_c}}U'^{\frac{1}{N_c}}w_r z_j\right)\prod_{k=1}^{N_f-N_c-1}\Gamma\left((pq)^{\frac{N_f-N_c}{2N_f}+x}U'^{-\frac{1}{N_f-N_c-1}}\sigma_{2,k}z_j \right)\nonumber\\
&\prod_{i=1}^{N_f-1}\Gamma\left((pq)^{\frac{N_f-N_c}{2N_f}-x}\tilde{t}_{N_f}^{\frac{1}{N_f-1}}\tau_i^{-1} z_j^{-1}\right)\prod_{a=1}^{N_c-1}\frac{dw_a}{2\pi \textup{i} w_a}\frac{dz_a}{2\pi \textup{i} z_a},
\end{align}

\noindent where $U'=\prod_{k=1}^{N_c+1}\tilde{s}_k$, $\sigma_{1,l}=U'^{-\frac{1}{N_c+1}}\tilde{s}_l$, $\sigma_{2,l}=U'^{\frac{1}{N_f-N_c-1}}\tilde{s}_{N_c+1+l}$ and $\tau_l=\tilde{t}_{N_f}^{\frac{1}{N_f-1}}\tilde{t}_l$. In complete analogy to the previous case, we may interprete this quantities, together with $\tilde{t}_{N_f}$, as fugacities corresponding to regular subgroups of the flavour symmetries. The structure is precisely the same, only the original factors are interchanged. We now have $\mathrm{SU(N_c+1)_{\sigma_1}\times SU(N_f-N_c-1)_{\sigma_2}\times U(1)_{\sigma}}\subset\mathrm{SU(N_f)_s}$ and $\mathrm{SU(N_f-1)_\tau\times U(1)_\tau}\subset \mathrm{SU(N_f)_t}$, as can be made apparent from the decomposition of the index of electric SQCD:

\begin{table}[t]
\centering
\begin{adjustbox}{max width=\textwidth}
\begin{tabular}{lccccccccc}
\toprule
 &  $\mathrm{SU(N_c)_z}$  & $\mathrm{SU(N_c)_w}$ & $\mathrm{SU(N_f-N_c-1)_\tau}$ & $\mathrm{SU(N_c+1)_\tau}$ & $\mathrm{SU(N_f-1)_\sigma}$ & $\mathrm{U(1)_\tau}$ & $\mathrm{U(1)_\sigma}$ & $\mathrm{U(1)_B}$ & $\mathrm{U(1)_R}$\\
\toprule
$\mathbf{Q}_1$ & $f$ & $1$ & $1$ & $1$ & $\bar{f}$ & $\frac{1}{N_f-1}$ & $0$ & $-1$ & $\frac{N_f-N_c}{N_f}$\\
$\mathbf{\bar{Q}}_1$ & $f$ & $1$ & $f$ & $1$ & $1$ & $0$ & $-\frac{1}{N_f-N_c-1}$ & $1$ & $\frac{N_f-N_c}{N_f}$\\
$\mathbf{Q}_2$ & $1$ & $f$ & $1$ & $1$ & $1$ & $-\frac{1}{N_c}$ & $-\frac{N_c-1}{N_c}$ & $-N_c$ & $\frac{N_c^2-N_cN_f+2N_f-2}{Nf}$\\
$\mathbf{\bar{Q}}_2$ & $1$ & $\bar{f}$ & $1$ & $f$ & $1$ & $\frac{1}{N_c}$ & $-\frac{1}{N_c}+\frac{1}{N_c+1}$ & $0$ & $\frac{2}{N_f}$\\
$\mathbf{Q}_{12}$ & $f$ & $f$ & $1$ & $1$ & $1$ & $-\frac{1}{N_c}$ & $\frac{1}{N_c}$ & $-1$ & $\frac{N_f-N_c-2}{N_f}$\\
$\mathbf{M}$ & $1$ & $1$ & $1$ & $f$ & $1$ & $-1$ & $\frac{1}{N_c+1}$ & $0$ & $2\frac{N_f-N_c}{N_f}$\\
$\mathbf{B}_1$ & $1$ & $1$ & $1$ & $1$ & $1$ & $1$ & $-1$ & $-N_c$ & $2+\frac{N_c(N_c+2-N_f)}{N_f}$\\
$\mathbf{B}_2$ & $1$ & $1$ & $1$ & $\bar{f}$ & $1$ & $1$ & $\frac{N_c}{N_c+1}$ & $N_c$ & $\frac{N_c(N_f-N_c)}{N_f}$\\
$\mathbf{V}_1$ & $\mathrm{adj}$ & $1$ & $1$ & $1$ & $1$ & $0$ & $0$ & $0$ & $1$\\
$\mathbf{V}_2$ & $1$ & $\mathrm{adj}$ & $1$ & $1$ & $1$ & $0$ & $0$ & $0$ & $1$\\
\toprule
\end{tabular}
\end{adjustbox}
\caption{The matter content of the quiver gauge theory whose field content corresponds to the superconformal index $\mathcal{I}_{\mathrm{quiver},\mathcal{T}}$ given by Eq. (\ref{quiverindex2}).}
\label{quivertable2}
\end{table}

\begin{align}\label{subgroups2}
&\mathcal{I}_{\mathrm{eSQCD}}=\kappa_n\int_{\mathbb{T}_z^n}\prod_{1\leq j<k\leq N_c}
\frac{1}{\Gamma(z_jz_k^{-1},z_j^{-1}z_k)}\prod_{j=1}^{N_c}\prod_{l=1}^{N_c+1}\Gamma\left((pq)^{r_Q}U'^{\frac{1}{N_c+1}}\sigma_{1,l}z_j\right) \nonumber\\
 &\prod_{k=1}^{N_f-N_c-1}\Gamma\left((pq)^{r_Q}U'^{-\frac{1}{N_f-N_c-1}}\sigma_{2,k}z_j\right)\prod_{i=1}^{N_f-1}\Gamma\left((pq)^{r_{\tilde{Q}}}\tilde{t}_{N_f}^{\frac{1}{N_f-1}}\tau_i^{-1} z_j^{-1}\right)\Gamma\left((pq)^{r_{\tilde{Q}}}\tilde{t}_{N_f}^{-1}z_i^{-1}\right)\prod_{a=1}^{N_c-1}\frac{dz_a}{2\pi \textup{i} z_a}.
\end{align}

The fact that the indices of both quivers are equivalent to that of ordinary electric SQCD indicates a nontrivial RG flow connecting theories with field content as listed in Tables \ref{quivertable} and \ref{quivertable2}. One interpretation is that one of the gauge nodes of each theory s-confines, leaving only one nontrivial gauge group. This is supported by the very derivation of the identities leading to the equivalence of superconformal indices: as explained in Section 2.2, Eq. \eqref{recursion1} (and analogously  Eq. \eqref{recursion2}) is derived by applying Eq. \eqref{An-int} to one of the integrals corresponding to one of the gauge groups. But \eqref{An-int} itself is interpreted physically in terms of s-confinement, as the index on the left hand (electric) side contains an integral, while the one on the right hand side (magnetic) does not.     

However, the flavour symmetries of the quivers do not match with those of SQCD, but corresponds to regular subgroups, and this adds another nontrivial element to the physical interpretation. Such a mismatch may be explained by the presence of a nontrivial superpotential which breaks the original symmmetry group. One possibility is then that following the RG flow the theory eventually reaches a point where it s-confines, and flavour symmetry is enhanced such that one arrives at electric SQCD. It would be interesting to to study these theories in more detail to see whether the present quivers can be related to ones where the symmetry is not broken.


\subsection{Anomaly matching}

We have calculated the coefficients for all triangle anomaly diagrams corresponding to the quivers of Eqs. \eqref{quiverindex1} and \eqref{quiverindex2}, and all gauge anomalies vanish. Furthermore, we have compared global anomalies with those of the subgroups of SQCD made explicit in Eqs. (\ref{subgroups1}) and \eqref{subgroups2}, respectively. The results agree exactly for all coefficients. For the $\mathcal{S}$-quiver the nonvanishing ones are given by

\begin{align}
    \mathrm{SU(N_c+1)^3} & &-N_c & & \mathrm{SU(N_c+1)^2\times U(1)_\tau} & &-\frac{N_c}{N_c+1} \nonumber\\
    \mathrm{SU(N_f-N_c-1)^3} & &-N_c & & \mathrm{SU(N_c+1)^2\times U(1)_B}& &-N_c \nonumber\\
    \mathrm{SU(N_f-1)^3} & &N_c & & \mathrm{SU(N_c+1)^2\times U(1)_R} & &-\frac{N_c^2}{N_f} \nonumber\\
    \mathrm{SU(N_f-N_c-1)^2\times U(1)_\tau} & & \frac{N_c}{N_f-N_c-1} & & \mathrm{SU(N_f-1)^2\times U(1)_\sigma} & & \frac{N_c}{1-N_f} \nonumber\\
    \mathrm{SU(N_f-N_c-1)^2\times U(1)_B} & & -N_c & & \mathrm{SU(N_f-1)^2\times U(1)_B} & & N_c \nonumber\\
    \mathrm{SU(N_f-N_c-1)^2\times U(1)_R} & & -\frac{N_c^2}{N_f} & & \mathrm{SU(N_f-1)^2\times U(1)_R} & & -\frac{N_c^2}{N_f}\nonumber
\end{align}
\begin{align}\label{A1}
\mathrm{U(1)_\sigma^3} && N_c-\frac{N_c}{(N_f-1)^2} && \mathrm{U(1)_\tau^3} && \frac{N_c}{(N_c+1-N_f)^2}-\frac{N_c}{(N_c+1)^2} \nonumber\\
\mathrm{U(1)_\sigma^2\times U(1)_B} && \frac{N_c N_f}{N_f-1} && \mathrm{U(1)_\tau^2\times U(1)_B} && \frac{N_c N_f}{(N_c+1)(N_c+1-N_f)} \nonumber\\
\mathrm{U(1)_\sigma^2\times U(1)_R} && \frac{N_c^2}{1-N_f} && \mathrm{U(1)_\tau^2\times U(1)_R} && \frac{N_c^2}{(N_c+1)(N_c+1-N_f)} \nonumber\\
\mathrm{U(1)_B^2\times U(1)_R} && -2N_c^2 && \mathrm{U(1)_R^3} && N_c^2-\frac{2N_c^4}{N_f^2}-1 \nonumber\\
\mathrm{U(1)_R} && -N_c^2-1 && \; && \; \nonumber\\
\end{align}

\noindent while for the $\mathcal{T}$-quiver they read anomaly coefficients involving $\mathrm{U(1)_R}$ are identical, while all others differ by a minus sign.


\subsection{$\mathrm{N_f=N_c+2}$}

An interesting situation arises for $\mathrm{N_f=N_c+2}$, corresponding to $m=0$ in the language of section 2. In this case the $\mathrm{SU(N_f-N_c-1)}$ flavour symmetry becomes trivial for both quivers $\mathcal{S}$ and $\mathcal{T}$. In particular, there exists a dual description obtained from the right hand side of Eq. \eqref{enhancement}. This can again be understood as in terms s-confinement, as Eq. \eqref{enhancement} is obtained by applying Eq. \eqref{An-int} to Eq. \eqref{recursion1}, this time to the other integral as compared to what is used to prove Eq. \eqref{recursion1} itself.

We proceed with the interpretation of Eq. (\ref{enhancement}). The left-hand side is the superconformal index of $\mathcal{N}=1$ electric SQCD as described in section 3.2, with the additional constraint $N_f=N_c+2$, and as such its field content is given by Table \ref{electricmatter}. The right-hand side, however, is more complicated. There are in total eight chiral multiplets and four of them are gauge singlets. Furthermore, the flavour symmetries are different. The left-hand side has $\mathrm{SU(N_f)\times SU(N_f)\times U(1)_B}$ symmetry, while the right-hand side has $\mathrm{SU(N_f-1)\times SU(N_f-1)\times U(1)\times U(1)\times U(1)_B}$. To make this explicit, we write the superconformal index on the right-hand side of Eq. (\ref{enhancement}), which we denote $\mathcal{I}_{\mathrm{II}}$ to indicate that it is the second theory to arise from s-confinement of the original quiver, such that the field content can be read off directly:

\begin{align}
&\mathcal{I}_{\mathrm{II}}=\prod_{k=1}^{N_c-1}\Gamma\left((pq)^{\frac{2}{N_f}}\tilde{s}_{N_f}\tilde{t}_{N_f}^{\frac{1}{N_f-1}}\right)\Gamma\left((pq)^{\frac{N_c}{N_f}-N_c x}\tilde{t}_{N_f}^{1-\frac{1}{N_f-1}}\tau_k\right)\nonumber\\
&\times\Gamma\left((pq)^{\frac{2}{N_f}}\tilde{t}_{N_f}^{-1}\tilde{s}_{N_f}^{-\frac{1}{N_f-1}}\sigma_k\right)\Gamma\left((pq)^{\frac{N_c}{N_f}+N_c x}\tilde{s}_{N_f}^{\frac{1}{N_f-1}-1}\sigma_k^{-1}\right)\nonumber\\
&\times\kappa_n\int_{\mathbb{T}^n}\prod_{j=1}^{N_c}\prod_{l=1}^{N_f-1}\Gamma\left((pq)^{\frac{1}{N_f}}\tilde{s}_{N_f}^{-\frac{1}{N_c}}\tilde{t}_{N_f}^{\frac{1}{N_c+1}-\frac{1}{N_c}}\tau_l^{-1}z_j^{-1}\right)\Gamma\left((pq)^{\frac{1}{N_f}}\tilde{t}_{N_f}^{\frac{1}{N_c}}\tilde{s}_{N_f}^{\frac{1}{N_c}-\frac{1}{N_c+1}}\sigma_l z_j\right)\nonumber\\
&\times\Gamma\left((pq)^{\frac{1}{N_f}-N_c x}\tilde{s}_{N_f}^{\frac{N_c-1}{N_c}}\tilde{t}_{N_f}^{-\frac{1}{N_c}}z_j^{-1}\right)\Gamma\left((pq)^{\frac{1}{N_f}+N_c x}\tilde{s}_{N_f}^{\frac{1}{N_c}}\tilde{t}_{N_f}^{-\frac{N_c-1}{N_c}}\right)\prod_{i=1}^{N_c-1}\frac{dz_i}{2\pi \textup{i} z_i},
\end{align}

\begin{table}[t]
\centering
\begin{tabular}{lccccccc}
\toprule
 &  $\mathrm{SU(N_c)}$  & $\mathrm{SU(N_f-1)_\sigma}$ & $\mathrm{SU(N_f-1)_\tau}$ & $\mathrm{U(1)_\sigma}$ & $\mathrm{U(1)_\tau}$ & $\mathrm{U(1)_B}$ & $\mathrm{U(1)_R}$\\
\toprule
$\mathbf{Q}^i$ & $f$ & $f$ & $1$ & $\frac{1}{N_c}-\frac{1}{N_c+1}$ & $\frac{1}{N_c}$ & $0$ & $\frac{2}{N_c+2}$ \\
$\mathbf{Q}_i$ & $\bar{f}$ & $1$ & $\bar{f}$ & $-\frac{1}{N_c}$ & $\frac{1}{N_c+1}-\frac{1}{N_c}$ & $0$ & $\frac{2}{N_c+2}$ \\
$\mathbf{Q}^{N_c+2}$ & $f$ & $1$ & $1$ & $\frac{1}{N_c}$ & $-\frac{N_c-1}{N_c}$ & $N_c$ & $\frac{2}{N_c+2}$ \\
$\mathbf{Q}_{N_c+2}$ & $\bar{f}$ & $1$ & $1$ & $\frac{N_c-1}{N_c}$ & $-\frac{1}{N_c}$  & $-N_c$ & $\frac{2}{N_c+2}$ \\
$\mathbf{M}_1$ & $1$ & $1$ & $\bar{f}$ & $1$ & $\frac{1}{N_c+1}$ & $0$ & $\frac{4}{N_c+2}$ \\
$\mathbf{M}_2$ & $1$ & $f$ & $1$ & $-\frac{1}{N_c+1}$ & $-1$ & $0$ & $\frac{4}{N_c+2}$ \\
$\mathbf{B}_1$ & $1$ & $1$ & $f$ & $0$ & $1-\frac{1}{N_c+1}$ & $-N_c$ & $\frac{2N_c}{N_c+2}$ \\
$\mathbf{B}_2$ & $1$ & $\bar{f}$ & $1$ & $\frac{1}{N_c+1}-1$ & $0$ & $N_c$ & $\frac{2N_c}{N_c+2}$ \\
$\mathbf{V}''$ & $\mathrm{adj}$ & $1$ & $1$ & $0$ & $0$ & $0$ & $1$ \\
\toprule
\end{tabular}
\caption{The matter content of the theory corresponding to the SCI on the right-hand side of Eq. (\ref{enhancement}). Note the relation $N_f=N_c+2$.}
\label{enhmatter}
\end{table}

\noindent where $\sigma_l=\tilde{s}_{N_f}^{\frac{1}{N_f-1}}\tilde{s}_l$ and $\tau_l=\tilde{t}_{N_f}^{\frac{1}{N_f-1}}\tilde{t}_l$ are the fugacities of the $\mathrm{SU(N_f-1)}$ factors, with $\prod_{l=1}^{N_f-1}\sigma_l=\prod_{l=1}^{N_f-1}\tau_l=1$, while $\tilde{s}_{N_f}$ and $\tilde{t}_{N_f}$ are the fugacities of the $\mathrm{U(1)}$ factors, and again $x$ labels the baryonic symmetry. One can rewrite the index on the left-hand side of Eq. (\ref{enhancement}) in terms of these fugacities, which shows that they correspond to actual subgroups of the $\mathrm{SU(N_f)_s\times SU(N_f)_t}$ flavour symmetries. The explicit expression is given by

\begin{align}
\mathcal{I}_{SQCD}&=\kappa_n\int_{\mathbb{T}^n}\prod_{j=1}^{N_c}\prod_{l=1}^{N_f-1}\Gamma\left((pq)^{N_f-x}\tilde{t}_{N_f}^{\frac{1}{N_f-1}}\tau_l^{-1}z_j^{-1}\right)\Gamma\left((pq)^{N_f+x}\tilde{s}_{N_f}^{-\frac{1}{N_f-1}}\sigma_l z_j\right)\nonumber\\
&\times\Gamma\left((pq)^{N_f-x}\tilde{t}_{N_f}^{-1}z_j^{-1}\right)\Gamma\left((pq)^{N_f+x}\tilde{s}_{N_f}z_j\right)\prod_{i=1}^{N_c-1}\frac{dz_i}{2\pi \textup{i} z_i}.
\end{align}

We have again calculated the coefficients for all global symmetries and found them to match with those of regular subgroups of SQCD. The nonvanishing coefficients are given by

\begin{align}
    \mathrm{SU(N_f-1)_\tau^3} & &-N_c & & \mathrm{SU(N_f-1)_\sigma^3 }& &N_c \nonumber\\
    \mathrm{SU(N_f-1)_\tau^2\times U(1)_\tau} & &\frac{N_c}{N_c+1} & & \mathrm{SU(N_f-1)_\sigma^2\times U(1)_\sigma}& &-\frac{N_c}{N_c+1}\nonumber\\
    \mathrm{SU(N_f-1)_\tau^2\times U(1)_B} & &-N_c & & \mathrm{SU(N_f-1)_\sigma^2\times U(1)_B }& &N_c \nonumber\\
    \mathrm{U(1)_\tau^3} & & -\frac{N_c^2N_f}{(N_c+1)^2} & &  \mathrm{U(1)_\sigma^3} & & \frac{N_c^2N_f}{(N_c+1)^2} \nonumber\\
    \mathrm{U(1)_\tau^2\times U(1)_B} & & -\frac{N_cN_f}{N_c+1}& & \mathrm{U(1)_\sigma^2\times U(1)_B} & & \frac{N_cN_f}{N_c+1} \nonumber\\
    \mathrm{SU(N_f-1)_\tau^2\times U(1)_R} & & -\frac{N_c^2}{N_f} & & \mathrm{SU(N_f-1)_\sigma^2\times U(1)_R} & & -\frac{N_c^2}{N_f}\nonumber\\
    \mathrm{U(1)_\tau^2\times U(1)_R} & & -\frac{N_c^2}{N_c+1} & &\mathrm{U(1)_\sigma^2\times U(1)_R} & &  -\frac{N_c^2}{N_c+1}\nonumber\\
    \mathrm{U(1)_B^2\times U(1)_R} & & -2N_c^2 & &\mathrm{U(1)_R} & &  -1-N_c^2\nonumber
\end{align}
\begin{align}
    \mathrm{U(1)_R^3 }& & -\frac{4+N_c(4+N_c(-3+(N_c-4)N_c))}{N_f^2}, & & & &
\end{align}

\noindent where we have used $N_f=N_c+2$ to simplify otherwise lengthy expressions.

This result again hints towards the presence of a superpotential deformation of the original theory such that flavour symmetry is broken down to the subgroup $SU(N_f-1)\times SU(N_f-1)\times U(1)\times U(1)$. We leave a detailed study of these deformations for the future. For now, let us summarise that we have, in the case of $\mathrm{N_f=N_c+2}$:

\begin{itemize}
 \item We have a set of four distinct theories, two of them quivers with a product gauge group, whose superconformal indices match due to the Bailey recursions (\eqref{recursion1} and \eqref{recursion2}) and the $\mathrm{A_n}$ integral \eqref{An-int}:
 
 \begin{equation}
  \mathcal{I}_{\mathrm{eSQCD}}=\mathcal{I}_{\mathrm{II}}=\mathcal{I}_{\mathrm{quiver},\mathcal{S}}=\mathcal{I}_{\mathrm{quiver},\mathcal{T}}.
 \end{equation}
 
 \item The field content of the theories corresponding to $\mathcal{I}_{\mathrm{II}}$, $\mathcal{I}_{\mathrm{quiver},\mathcal{S}}$, and $\mathcal{I}_{\mathrm{quiver},\mathcal{T}}$ transforms in regular subgroups of the flavour symmetries of electric SQCD. This suggests the presence of a superpotential breaking the larger symmetry.

 \item From the derivation of the corresponding indices it seems that the two non-quiver theories arise as s-confining descriptions of the two quivers. One may switch to the s-confining description of either of both nodes, and obtain one of the two theories accordingly.
 
 \item A possible explanation for the equivalence of superconformal indices is of course the presence of a nontrivial RG flow connecting all of them. It is not clear, however, that all four theories are equivalent descriptions of a theory at one and \emph{the same} fixed point, i.e. it is not guaranteed that the theory corresponding to $\mathcal{I}_{\mathrm{IR}}$ is actually dual to electric SQCD.
 
 \item At this point we should mention that the Bailey recursion may also be applied to the magnetic side of original Seiberg duality, as the corresponding index is also given by an $\mathrm{A_n}$ integral, although with different parameters, as it is equivalent to the right hand side of Eq. \eqref{rains2}. This leads to an additional set of theories connected to those generated by the electric side.
 
\end{itemize}

\subsection{$\mathrm{N_f>N_c+2}$}

For $\mathrm{N_f>N_c+2}$ we are presented with the scenario that the sector of the quiver corresponding to the $z_j$ fugacities does not s-confine, but instead has a magnetic dual description in terms of nontrivial gauge group, given by $\mathrm{SU(N_f-N_c-1})$. This is reflected in the double integral on the right hand side of Eq. \eqref{BS}, which we rewrite in terms of the variables used above as

\begin{table}[t]
\centering
\begin{adjustbox}{max width=\textwidth}
\begin{tabular}{lccccccccc}
\toprule
 &  $\mathrm{SU(N_f-N_c-1)_z}$  & $\mathrm{SU(N_c)_w}$ & $\mathrm{SU(N_f-N_c-1)_\tau}$ & $\mathrm{SU(N_c+1)_\tau}$ & $\mathrm{SU(N_f-1)_\sigma}$ & $\mathrm{U(1)_\sigma}$ & $\mathrm{U(1)_\tau}$ & $\mathrm{U(1)_B}$ & $\mathrm{U(1)_R}$\\
\toprule
$\mathbf{q}_1$ & $f$ & $1$ & $1$ & $1$ & $\bar{f}$ & $\frac{1}{N_f-1}-\frac{1}{N_f-N_c-1}$ & $0$ & $\frac{N_c}{N_f-N_c-1}$ & $\frac{N_c(N_f-N_c-2)}{N_f(N_f-N_c-1}$\\
$\mathbf{\bar{q}}_1$ & $\bar{f}$ & $1$ & $f$ & $1$ & $1$ & $\frac{1}{N_f-N_c-1}$ & $-\frac{1}{N_f-N_c-1}$ & $-\frac{N_c}{N_f-N_c-1}$ & $\frac{N_c(N_f-N_c-2)}{N_f(N_f-N_c-1)}$\\
$\mathbf{q}_2$ & $1$ & $f$ & $1$ & $1$ & $1$ & $\frac{1}{N_c}$ & $\frac{N_c-1}{N_c}$ & $N_c$ & $\frac{N_c^2-N_cN_f+2N_f-2}{Nf}$\\
$\mathbf{\bar{q}}_2$ & $1$ & $\bar{f}$ & $1$ & $\bar{f}$ & $1$ & $-\frac{1}{N_c}$ & $\frac{1}{N_c}-\frac{1}{N_c+1}$ & $0$ & $\frac{2}{N_f}$\\
$\mathbf{q}_{12}$ & $\bar{f}$ & $\bar{f}$ & $1$ & $1$ & $1$ & $\frac{1}{N_f-N_c-1}-\frac{1}{N_c}$ & $\frac{1}{N_c}$ & $-\frac{N_c}{N_f-N_c-1}$ & $\frac{N_c^2+4Nc+2-(N_c+2)N_f}{N_f(N_c+1-N_f)}$\\
$\mathbf{M}$ & $1$ & $1$ & $1$ & $\bar{f}$ & $1$ & $1$ & $-\frac{1}{N_c+1}$ & $0$ & $2\frac{N_f-N_c}{N_f}$\\
$\mathbf{M'}$ & $1$ & $1$ & $\bar{f}$ & $1$ & $f$ & $-\frac{1}{N_f-1}$ & $\frac{1}{N_f-N_c-1}$ & $0$ & $2\frac{N_f-N_c}{N_f}$\\
$\mathbf{B}_1$ & $1$ & $1$ & $1$ & $1$ & $1$ & $-1$ & $1$ & $N_c$ & $2+\frac{N_c(N_c+2-N_f)}{N_f}$\\
$\mathbf{B}_2$ & $1$ & $1$ & $1$ & $f$ & $1$ & $1$ & $-\frac{N_c}{N_c+1}$ & $-N_c$ & $\frac{N_c(N_f-N_c)}{N_f}$\\
$\mathbf{V}_1$ & $\mathrm{adj}$ & $1$ & $1$ & $1$ & $1$ & $0$ & $0$ & $0$ & $1$\\
$\mathbf{V}_2$ & $1$ & $\mathrm{adj}$ & $1$ & $1$ & $1$ & $0$ & $0$ & $0$ & $1$\\
\toprule
\end{tabular}
\end{adjustbox}
\caption{The matter content of the quiver gauge theory whose field content corresponds to the superconformal index $\mathcal{I}_{\mathrm{quiver},\mathcal{S}}$ given by Eq. (\ref{quiverindex3}).}
\label{quivertable3}
\end{table}

\begin{align}\label{quiverindex3}
&\mathcal{I}_{\mathrm{quiver},\mathcal{S}}=\kappa_n \kappa_m \Gamma\left((pq)^{1+\frac{N_c}{2N_f}(2+N_c-N_f)+N_c x}\tilde{s}_{N_f}^{-1}U\right)\nonumber\\
&\times\prod_{l=1}^{N_c+1}\Gamma\left((pq)^{\frac{N_f-N_c}{N_f}}\tilde{s}_{N_f}U^{-\frac{1}{N_c+1}}\tau_{1,l}^{-1}\right)\Gamma\left((pq)^{\frac{N_c(N_f-N_c)}{2N_f}-N_c x}U^{-\frac{N_c}{N_c+1}}\tau_{1,l}\right)\nonumber\\
&\times\prod_{k=1}^{N_f-1}\prod_{l=1}^{N_f-N_c-1}\Gamma\left((pq)^{\frac{N_f-N_c}{N_c}}\tilde{s}_{N_f}^{-\frac{1}{N_f-1}}U^{\frac{1}{N_f-N_c-1}}\sigma_k\tau_{2,l}^{-1}\right)\nonumber\\
&\times\int_{\mathbb{T}_w^n}\int_{\mathbb{T}_z^m}\prod_{1\leq j<k\leq N_c}
\frac{1}{\Gamma(w_jw_k^{-1},w_j^{-1}w_k;p,q)}\prod_{1\leq j<k\leq N_c-N_f-1}
\frac{1}{\Gamma(z_jz_k^{-1},z_j^{-1}z_k;p,q)}\nonumber
\end{align}
\begin{align}
&\times\prod_{j=1}^{N_c}\Gamma\left((pq)^{\frac{(N_c^2-N_c N_f+2N_f-2)}{2N_f}+N_c x}\tilde{s}_{N_f}^{\frac{1}{N_c}}U^{\frac{N_c-1}{N_c}}w_j\right)\prod_{l=1}^{N_c+1}\Gamma\left((pq)^{\frac{1}{N_f}}\tilde{s}_{N_f}^{-1}U^{\frac{1}{N_c}-\frac{1}{N_c+1}}\tau_{1,l}^{-1}w_j^{-1}\right)\nonumber\\
&\times\prod_{i=1}^{N_f-1}\Gamma\left((pq)^{\frac{N_f-N_c-1)}{N_f}}\tilde{s}_{N_f}^{\frac{1}{N_f-1}-\frac{1}{N_f-N_c-1}}U^{-\frac{1}{N_c}}\sigma_i w_j\right)\nonumber \\
&\times\prod_{k=1}^{N_f-N_c-1}\Gamma\left((pq)^{\frac{N_c^2+4N_c+2-(N_c+2)N_f}{2N_f(N_c+1-N_f}-x\frac{N_c}{N_f-N_c-1}}\tilde{s}_{N_f}^{\frac{1}{N_f-N_c-1}-\frac{1}{N_c}}U^{\frac{1}{N_c}}w_j^{-1} z_k^{-1}\right)\nonumber\\
&\times\prod_{r=1}^{N_f-N_c-1}\Gamma\left((pq)^{\frac{N_c(N_f-N_c-2)}{2N_f(N_f-N_c-1)}-x\frac{N_c}{N_f-N_c-1}}\tilde{s}_{N_f}^{\frac{1}{N_f-N_c-1}}U^{-\frac{1}{N_f-N_c-1}}\tau_{2,k} z_j^{-1}\right)\nonumber\\
&\times\Gamma\left((pq)^{\frac{N_c(N_f-N_c-2)}{2N_f(N_f-N_c-1)}+x\frac{N_c}{N_f-N_c-1}}\tilde{s}_{N_f}^{\frac{1}{N_f-1}-\frac{1}{N_f-N_c-1}}\sigma_i^{-1}z_k\right)\prod_{a=1}^{N_c-1}\frac{dw_a}{2\pi \textup{i} w_a}\frac{dz_a}{2\pi \textup{i} z_a}.
\end{align}

\noindent We summarize the field content of the corresponding theory in Table \ref{quivertable3}. Again the fields are organized in representations of regular subgroups of the flavour symmetries of electric SQCD. The field content is of course similar, apart from different charge assignments and the presence of two additional fields: one mesonic gauge singlet and one field that is charged under the gauge symmetry corresponding to the fugacity $w_i$. This is due to the involvement of the magnetic description of a sector of the quiver. Anomaly matching may be verified in an analogous manner as above.

\section{A Bailey tree of quivers}

\subsection{An example: $\mathrm{N_f=N_c+3}$}

The principle of interpreting integral identities in terms of equivalences of superconformal indices that provide evidence for new dualities may be generalised to arbitrarily complicated identities that may be formed from the building blocks as explained in Section 2. Combining the operations \eqref{r1}, \eqref{r2}, and \eqref{crec} can, depending on the choice of $n$ and $m$, lead to very large families of related physical theories. In the following, we will build the corresponding Bailey tree for $\mathrm{N_f=N_c+3}$, i.e. we will consider possible ways of rewriting $I_n^{(2)}$, making use of the aforementioned operations. For simplicity, we will consider the action on the first $n+1$ parameters, and omit the arguments of corresponding $\mathcal{S}$ and $\mathcal{T}$ operators. We then obtain

\begin{align}
 I_n^{(2)}=\mathcal{S}_n^1I_n^{(1)}=\mathcal{T}^1_n I_n^{(1)}=\mathcal{S}^1_n\mathcal{S}^0_n I^{(0)}_n=\mathcal{T}^1_n\mathcal{T}^0_n I^{(0)}_n=\mathcal{S}^1_n\mathcal{T}^0_n I^{(0)}_n=\mathcal{T}^1_n\mathcal{S}^0_n I^{(0)}_n.
\end{align}

\noindent Seiberg duality now generically leads to many more relations. For simplicity, we will consider $N_c=2$, i.e. $n=1$. This leads to

\begin{align}
 I_1^{(2)}=\mathcal{S}_1^1I_1^{(1)}=\mathcal{T}^1_1 I_1^{(1)}=\mathcal{S}^1_1\mathcal{S}^0_1 I^{(0)}_1=\mathcal{T}^1_1\mathcal{T}^0_1 I^{(0)}_1=\mathcal{S}^1_1\mathcal{T}^0_1 I^{(0)}_1=\mathcal{T}^1_1\mathcal{S}^0_1 I^{(0)}_1,
\end{align}

\noindent with the additional equalities

\begin{equation}
 \mathcal{S}_1^1I_1^{(1)}=\mathcal{S}_1^1\mathcal{C}^1_1I_1^{(1)}=\mathcal{S}_1^1\mathcal{C}^1_1\mathcal{S}^0_1 I^{(0)}_1=\mathcal{S}_1^1\mathcal{C}^1_1\mathcal{T}^0_1 I^{(0)}_1=\mathcal{S}_1^1\mathcal{C}^1_1\mathcal{S}^0_1 \mathcal{C}^0_1I^{(1)}_0=\mathcal{S}_1^1\mathcal{C}^1_1\mathcal{T}^0_1 \mathcal{C}^0_1I^{(1)}_0,
\end{equation}

\begin{equation}
 \mathcal{T}_1^1I_1^{(1)}=\mathcal{T}_1^1\mathcal{C}^1_1I_1^{(1)}=\mathcal{T}_1^1\mathcal{C}^1_1\mathcal{S}^0_1 I^{(0)}_1=\mathcal{T}_1^1\mathcal{C}^1_1\mathcal{T}^0_1 I^{(0)}_1=\mathcal{T}_1^1\mathcal{C}^1_1\mathcal{S}^0_1 \mathcal{C}^0_1I^{(1)}_0=\mathcal{T}_1^1\mathcal{C}^1_1\mathcal{T}^0_1 \mathcal{C}^0_1I^{(1)}_0,
\end{equation}

\begin{equation}
 \mathcal{S}^1_1\mathcal{S}^0_1 I^{(0)}_1=\mathcal{S}^1_1\mathcal{S}^0_1 \mathcal{C}^0_1 I^{(1)}_0,
\end{equation}

\begin{equation}
 \mathcal{S}^1_1\mathcal{T}^0_1 I^{(0)}_1=\mathcal{S}^1_1\mathcal{T}^0_1 \mathcal{C}^0_1 I^{(1)}_0,
\end{equation}

\begin{equation}
 \mathcal{T}^1_1\mathcal{S}^0_1 I^{(0)}_1=\mathcal{T}^1_1\mathcal{S}^0_1 \mathcal{C}^0_1 I^{(1)}_0,
\end{equation}

\noindent

\begin{equation}
 \mathcal{T}^1_1\mathcal{T}^0_1 I^{(0)}_1=\mathcal{T}^1_1\mathcal{T}^0_1 \mathcal{C}^0_1 I^{(1)}_0.
\end{equation}

\noindent Thus, the Bailey lemma implies that there are (at least, keep in mind other permutations of parameters) 20 distinct theory connected to electric SQCD, and a similar construction can be done for the magnetic side. The form the index of a quiver arising from a mixture of $\mathcal{S}$ and $\mathcal{T}$ operations is given for example by Eq. \eqref{snowflake}, with others corresponding to similar expressions. For brevity, we will refrain from listing the field content of all these theories explicitly, but want to point out that the physical mechanisms seem to be the same as indicated for the simpler examples of the previous section.   

\subsection{Extending the web: $\mathrm{SU}\leftrightarrow \mathrm{SU}$ dualities}

As pointed out in \cite{SV1}, there exists another set of dualities derived from mixed $\mathrm{SU}\leftrightarrow \mathrm{SU}$ transformations described in \cite{rains}. In the language of the present article, the transformations act on the $A_n$ integral \eqref{An-int-def} with $m=1$ and are given by

\begin{align}\label{SUSU}
&I_n^{(1)}(s_1,\dots,s_{n+3};t_1,\dots t_{n+3})=\nonumber\\
&\prod_{1\leq i<K,K\leq j\leq n+3}
\Gamma(s_i t_j,s_j t_i,s_i S/s_j,t_j T/t_i)I_n^{(1)}(s'_1,\dots,s'_{n+3};t'_1,\dots t'_{n+3}),
\end{align}

\noindent with

\begin{align}
s'_l=(S/T)^{\frac{n+1-K}{2(n+1)}}(S_K/T_K)^{\frac{1}{n+1}}t_l,\;\;\;\;\;1\leq l<K+1,\nonumber\\
s'_l=(T/S)^{\frac{K}{2(n+1)}}(S_K/T_K)^{\frac{1}{n+1}}s_l,\;\;\;\;\; K+1\leq l\leq n+3,\nonumber\\
t'_l=(T/S)^{\frac{n+1-K}{2(n+1)}}(T_K/S_K)^{\frac{1}{n+1}}t_l,\;\;\;\;\;1\leq l<K+1,\nonumber\\
t'_l=(S/T)^{\frac{K}{2(n+1)}}(T_K/S_K)^{\frac{1}{n+1}}s_l,\;\;\;\;\; K+1\leq l\leq n+3,\nonumber\\
\end{align}

\noindent where $S=\prod_l^{n+3}s_l$, $T=\prod_l^{n+3}t_l$, $S_K=\prod_l^{K}s_l$ and $T_K=\prod_l^{K}t_l$. The flavour symmetry on the left-hand side of Eq. (\ref{SUSU}) is given by $\mathrm{SU(N_c+2)\times SU(N_c+2)}$, while that on the right-hand side depends on the discrete parameter $K$, with $0<K<N_c+2$ and is given by $\mathrm{SU(K)\times SU(N_c+2-K)\times U(1)\times SU(K)\times SU(N_c+2-K)\times U(1)}$. This is a direct product of subgroups of the original flavour symmetry on the left-hand side, and hence points towards a similar mechanism as the one discussed in the previous sections. Eq.(\ref{SUSU}) does not follow from the Bailey lemma, but is a consequence of the fact that the Seiberg-dual theory has the gauge group $SU(2)$ for which the
flavour group is enlarged to $SU(2(N_c+2))$. Because of that, after all possible permutation of corresponding
fugacities and another application of the Seiberg duality transformation we get the described 
$\mathrm{SU}\leftrightarrow \mathrm{SU}$ duality, and as such it should be viewed as an extension of the Bailey tree of dualities. The consequence for the duality web is that each factor in a superconformal index that contains $I_n^{(1)}(s_1,\dots,s_{n+3};t_1,\dots,t_{n+3})$ acts as a starting point for a new branch in the duality tree. Denoting schematically the operation of Eq. (\ref{SUSU}) by $I_n^{(1)}=X_n^1 I_n^{(1)}$, it is possible to systematically extend the duality tree along the lines of the previous section. The indices corresponding to these additional theories are all equal to each other, either as a consequence of the Bailey lemma or of ordinary Seiberg duality.

In fact, this extension of the duality tree contains an additional set of dual quivers, as can be seen from alternating application of Seiberg duality and "X-duality" to $I_1^{(1)}$:

\begin{equation}
I_1^{(1)}=C_1^1 I_1^{(1)}=C_1^1 X_1^1 I_1^{(1)}=C_1^1 X_1^1 C_1^1 I_1^{(1)} = C_1^1 X_1^1 C_1^1 X_1^1 I_1^{(1)}=\dots\; .
\end{equation}

\noindent As also discussed in \cite{SV1}, there exist mixed $A_n\leftrightarrow BC_n$ ($\mathrm{SU}\leftrightarrow \mathrm{SP}$) transformations that in principle extend the duality web as well. We leave this for future work, as the present article is limited to the $A_n$ root system.

\subsection{A set of fully $s$-confining quivers}

Along the lines of the previous sections it can easily be seen that for generic values of $m$, the elliptic hypergeometric integral on the left-hand side of Eq. (\ref{sconfquiver}), after dividing by all elliptic gamma functions not depending on the integration variables, is equivalent to the superconformal index of a specific class of quivers with $m$ nodes, each corresponding to the gauge group $\mathrm{SU(N_c)}$ with $\mathrm{n=N_c}$. For $m=1$, it reduces to the index of ordinary SQCD in the $s$-confining region of the parameter space, i.e. $\mathrm{N_f=N_c+1}$. The $M$-operators contain the contribution of the bifundamental fields and vector multiplets, while the $D$-operators encode the flavour symmetries. Going up one level,  i.e. $m\rightarrow m+1$, one of the flavour symmetries is turned into a gauge symmetry and additional flavour symmetries are added. An important observation is that no matter which value of $m$ is chosen, the right-hand side of Eq. (\ref{sconfquiver}) will not involve an integration, i.e. it will correspond to the index of an $s$-confining theory. In other words, the quivers described by this
equation will always be $s$-confining, independently of their length.

Let us illustrate this for the example of $m=2$. The superconformal index of the linear quiver is given by

\begin{align}
&\mathcal{I}=\kappa_{N_c}\int_{\mathbb{T}^{N_c}_w}\int_{\mathbb{T}^{N_c}_z}\prod_{1\leq j<k\leq N_c}
\frac{1}{\Gamma(z_jz_k^{-1},z_j^{-1}z_k)}\frac{1}{\Gamma(w_jw_k^{-1},w_j^{-1}w_k)}\nonumber\\
&\prod_{j=1}^{N_c}\prod_{l=1}^{N_c+1}\Gamma(t_lz_j) \prod_{k=1}^{N_c} \Gamma(s_1w_kz_j^{-1})\prod_{i=1}^{2}\Gamma(\sigma_iw_l)\prod_{r=1}^{N_c+2}\Gamma(\tau_rw_l^{-1})\prod_{a=1}^{N_c-1}\frac{dz_a}{2\pi \textup{i} z_a}\frac{dw_a}{2\pi \textup{i} w_a},
\end{align}

\noindent with $z\equiv z^{(1)}$, $w\equiv z^{(2)}$. The parameters $t_l=tx_l^{-1}$ (for $l=1\dots N_c$) and $t_{N_c+1}=\sqrt{pq}(ts_1)^{-\frac{N_c+1}{2}}u_1^{-1}$ correspond to an $\mathrm{SU(N_c+1)}$ flavour group, the parameter $s_1$ corresponds to a $\mathrm{U(1)}$, the parameters $\sigma_1=\sqrt{pq}t^{\frac{N_c+1}{2}}s_1^{-\frac{N_c-1}{2}}u_1^{-1}$ and $\sigma_2=\sqrt{pq}(ts_1s_2)^{-\frac{N_c+1}{2}}u_2^{-1}$ contain the contributions of an $\mathrm{SU(2)}$ group, and finally, the parameters $\tau_l$ (for $l=1\dots N_c$), $\tau_{N_c+1}=\sqrt{pq}t^{\frac{N_c+1}{2}}s_1^{\frac{N_c-1}{2}}u_1$ and $\tau_{N_c+2}=\sqrt{pq}(ts_1s_2)^{-\frac{N_c+1}{2}}u_2$ correspond to an $\mathrm{SU(N_c+2)}$ group. It is again straightforward to see that the quiver is free of gauge anomalies. Summing the number of flavours of groups adjacent to a given gauge node and weighting them by the type of representation ($1$ for fundamental, $-1$ for antifundamental) gives zero, a condition for freedom from gauge
anomalies. What we also see is that the number of flavours attached to a given gauge node, again weighted by the type of the representation, is given by $\mathrm{N_c+1}$. This is true for both nodes, and the quiver fully $s$-confines, in agreement with the statement above.

Generalizing this idea to $m>2$ is a straightforward exercise. At each level, we get one more gauge group $\mathrm{SU(N_c)}$, two additional $\mathrm{SU(2)}$ factors and new U(1) factors. The $s$-confining behaviour of the theory however does not change.

\section{Conclusion}

In this article, we have studied the Bailey lemma for elliptic hypergeometric integrals on the $\mathrm{A_n}$ root system and  integral identities generated by that. Interpreting these integrals as superconformal indices of four-dimensonal $\mathcal{N}=1$ supersymmetric gauge theories, the resulting Bailey tree can be viewed as evidence for a network of quiver gauge theories connected by nontrivial RG flows and involving phenomena such as s-confinement and flavour symmetry breaking/enhancement. We have found further evidence by explicitly confirming the 't Hooft anomaly matching condition for some simple examples. We have also shown that the duality web is extended further by including ordinary Seiberg duality and another related set of dualities, both of which do not follow directly from the Bailey lemma.

It is worth pointing out that the duality relations obtained from the Bailey lemma hold for arbitrarily large values of the parameter $m$, i.e. for a large number of flavours. There is no obstruction to carrying this beyond the conformal window. This is reminiscent of the dualities studied in \cite{conformal}, which similarly were not limited to this regime. Remarkably, this phenomenon was correlated with the presence of small R-charges. This is also what happens in the case of the quivers discussed in this article: R-charges are inversely proportional to the number of flavours, and a large parameter $m$ leads to small values. 

There are several directions in which one could expand on our work. An important issue would be a systematic study of superpotential deformations underlying our network of theories. Another would be to construct the Bailey tree for integrals on different root systems. It is likely that in the simplest cases the corresponding physical theories will again correspond to $\mathcal{N}=1$ theories, but this time with different gauge groups. Furthermore it would also be interesting to see if one could apply the Bailey lemma to theories with extended supersymmetry. Another possibility would be to see whether the Bailey lemma can be applied not only to the ordinary superconformal index on $\mathrm{S^1\times S^3}$, but also to indices arising from other geometries such as the lens space \cite{BNY}. Elliptic hypergeometric integrals related to the latter were studied in \cite{kels,rare}. Furthermore, it would be interesting to see whether the theories arising from the Bailey lemma are in any way related to other examples where nontrivial flavour symmetry breaking/enhancement or s-confining phenomena appear \cite{Lillard,ASS,RSZ}.

\medskip
We would like to thank I\~{n}aki Garcia-Etxebarria and Timm Wrase for helpful
discussions. F.B. was supported by the Austrian Science Fund FWF, project
no.  P26366, and the FWF doctoral program Particles \& Interactions,
project no. W1252. Results of Section 2 have been worked out within the
RFBR grant no. 16-01-00562. V.S. is supported in part by Laboratory of
Mirror Symmetry NRU HSE, RF government grant, ag. no. 14.641.31.0001.

\end{document}